\documentclass[%
reprint,
superscriptaddress,
nofootinbib,
longbibliography,
noeprint,
amsmath,amssymb,
aps,pre
]{revtex4-2}
\usepackage{amsthm}

\usepackage{graphicx}% Include figure files
\usepackage{physics}
\usepackage{xcolor}
\graphicspath{{figures/},{pictures/},{fractals/}}
\usepackage[autostyle, english = american]{csquotes}
\usepackage{chngcntr}
\usepackage[ruled,vlined, linesnumbered]{algorithm2e}
\usepackage{tabularx}
\usepackage{cellspace}
\usepackage{bm}  % provides \boldsymbol
\renewcommand{\vec}[1]{\boldsymbol{\mathbf{#1}}}
\MakeOuterQuote{"}
\usepackage[caption=false,justification=centerlast]{subfig}
\definecolor{darkGreen}{RGB}{0,110,0}
\definecolor{darkBlue}{RGB}{0,0,130}

\usepackage[colorlinks,citecolor=darkGreen,linkcolor=darkBlue,urlcolor=blue,hyperindex]{hyperref}

\def\equationautorefname~#1\null{Eq. #1\null}
\def\algorithmautorefname~#1\null{Alg. #1\null}
\def\linerefname~#1\null{line #1\null}

\newcommand{\appref}[1]{\hyperref[#1]{App.~\ref*{#1}}}

\newcommand{\comment}[1]{}

\begin{document}
\date{\today} % delete this line to display the current date

\title{Self-dual quasiperiodic percolation}

\author{Grace M. Sommers}
\affiliation{Department of Physics, Princeton University, Princeton, NJ 08544, USA}

\author{Michael J. Gullans}
\affiliation{Joint Center for Quantum Information and Computer Science, NIST/University of Maryland, College Park, Maryland 20742, USA}

\author{David A. Huse}
\affiliation{Department of Physics, Princeton University, Princeton, NJ 08544, USA}
\affiliation{Institute for Advanced Study, Princeton, NJ 08540, USA}

\begin{abstract}
How does the percolation transition behave in the absence of quenched randomness? To address this question,
we study two nonrandom self-dual quasiperiodic models of square-lattice bond percolation. In both models, the critical point has emergent discrete scale invariance, but none of the additional emergent conformal symmetry of critical random percolation. From the discrete sequences of critical clusters, we find fractal dimensions of $D_f=1.911943(1)$ and $D_f=1.707234(40)$ for the two models, significantly different from $D_f=91/48=1.89583...$ of random percolation. The critical exponents $\nu$, determined through a numerical study of cluster sizes and wrapping probabilities on a torus, are also well below the $\nu=4/3$ of random percolation. While these new models do not appear to belong to a universality class, they demonstrate how the removal of randomness can fundamentally change the critical behavior.
\end{abstract}

\maketitle
\section{Introduction}
Random percolation is a foundational topic in statistical physics with many practical applications across engineering and science, as well as the subject of a series of exact results. These include the mapping between percolation and the $Q$-state Potts model with $Q\rightarrow 1$~\cite{Kasteleyn1969,Wu1978}, conformal invariance at the critical point~\cite{Cardy1992,Langlands1994,Smirnov2001,Duminil-Copin2020}, and a proven percolation threshold of $p=1/2$ for square lattice bond percolation~\cite{Kesten1980}.  Random percolation models are constructed by occupying bonds or sites of a lattice according to a probability distribution.  One then asks if an infinite connected cluster is present in the limit of an infinite lattice. 
%\gs{Is this an accurate statement? There are random percolation models with long range bond correlations also, so not just Gaussian...} 

In this paper, we study models in which the pattern of which bonds are occupied is deterministic and quasiperiodic rather than random, {finding critical behavior quite different from random percolation.} %which opens the door to a different universality class for the percolation phase transition. \gs{But we do not find universality, so should I say this a bit differently?} 
One application of these new classes of quasiperiodic models, and indeed our motivation for studying this question in the first place, is to measurement-induced phase transitions in quantum circuits~\cite{Li2018,Skinner2018,Li2019,Chan2019,Gullans2020}. In the minimal cut picture, a random circuit maps onto a percolation instance, with vertices corresponding to unitary gates and edges corresponding to qubits with some probability $p$ of being measured. This mapping describes the transition in the zeroth Renyi entropy for Haar-random circuits~\cite{Skinner2018}, whereas higher order Renyi entropies exhibit a transition at a much lower measurement rate that only belongs to the percolation universality class in the limit of infinite onsite Hilbert-space dimension~\cite{Jian2020,Bao2020}. These circuits can be designed without any randomness in the space-time pattern of gates and measurements. While the Born probabilities of the measurement outcomes will in general produce randomness, in Clifford circuits these random outcomes only induce signs on the stabilizer operators which do not affect the local order parameter that probes the transition~\cite{Gullans2019,Li2019}. The measurement-induced phase transition in a particular nonrandom circuit model---a Floquet Clifford circuit with quasiperiodic measurement locations---is studied in Ref.~\cite{Li2019}, but the quasiperiodic construction differs substantially from the models introduced in this paper. Our models are also quite different from the only other quasiperiodic percolation models in the literature that we are aware of, namely the continuum percolation models studied in Ref.~\cite{Chernikov1994}.

We define two classes of self-dual square lattice quasiperiodic bond percolation models. In the "checkerboard model," the quasiperiodic pattern of occupied bonds in square lattice $\mathcal{L}$ is specified by another square lattice $\mathcal{L'}$ that is rotated with respect to the $\mathcal{L}$ and divided into sublattices in a checkerboard fashion. In the "counter-rotated model," the quasiperiodic pattern is specified by a pair of lattices $\mathcal{L'}_A, \mathcal{L'}_B$, that are counter-rotated with respect to $\mathcal{L}$, thus giving the model %additional horizontal and vertical 
{reflection symmetries that the checkerboard model does not have.} These models both produce a deterministic and quasiperiodic ranking of all the bonds in the lattice $\mathcal{L}$.  The fraction $n$ of bonds occupied is tuned by occupying all bonds below a given level in the ranked list of all bonds. 

The models are self-dual by design, so the percolation transition is at $n_c=1/2$, where $n$ is the fraction of occupied bonds. For both models, the critical point has a discrete self-similarity, which allows us to obtain very accurate estimates of the fractal dimension $D_f$ and hull exponent $D_h$ of the infinite critical percolating cluster, both of which differ from the exponents $D_f=91/48=1.89583 ...$, $D_h=7/4$ of two-dimensional critical random percolation. No continuous scale invariance or full rotational symmetry emerges at the critical point, so it does not have conformal invariance. 

Our estimates of the correlation length exponent $\nu$ are much less precise, but well below the $\nu=4/3$ of random percolation. Also in contrast to random percolation, two-fold anisotropy in the checkerboard model is relevant:  When we let the occupations of the bonds in the two orientations differ, this opens up an apparently rich phase diagram that includes intermediate phases that percolate only along one of the two directions.
%, leading us to propose a 2d phase diagram parameterized by the occupation rates of opposite parity bonds.

Thus our models clearly lie outside the universality class for two-dimensional random percolation. The two main models studied here have different critical exponents but also different underlying symmetry. To probe universality, we further modify the counter-rotated model via a tunable parameter that preserves the underlying symmetry, and find that this modification changes the distribution and fractal dimension of critical clusters. This indicates an absence of universality, although it does not rule out the {possible existence of genuine universality classes for other nonrandom percolation models.}  We leave this and many other interesting questions unanswered, including the relation between these models and the continuum percolation models of Ref.~\cite{Chernikov1994}. %the exploration of other possible quasiperiodic percolation models, with or without self-duality.

{We report more results for the ``checkerboard'' model, because it is the model that we investigated first.  The ``counter-rotated'' models may have equally rich and detailed behavior, but once it seemed clear to us that we are not exploring a single new universality class, but instead multiple different cases (or perhaps classes), it seemed less interesting to explore a similar level of detail for all the models.}

This paper proceeds as follows. In~\autoref{sect:model}, we define the models, and specialize to a "maximally incommensurate" choice of $\mathcal{L'}$. The methods for probing the percolation transition on finite systems and nominally infinite systems are described in~\autoref{sect:methods}. Moving to the results, we obtain estimates of $D_f$ and $D_h$ for the checkerboard and counter-rotated models in~\autoref{sect:fractal}, providing evidence for the lack of universality. The critical exponent $\nu$ for both models is estimated in~\autoref{sect:nu}. %Since the behavior is model-specific, 
In~\autoref{sect:anisotropy} we focus on the checkerboard model, showing that two-fold anisotropy is relevant. Finally, we discuss these results and outstanding questions for future work in~\autoref{sect:discuss}.

\section{Model Details}
\label{sect:model}
In order to define a deterministic and quasiperiodic ranking of the bonds of square lattice $\mathcal{L}$, we need to answer two basic design questions. First is the choice of lattice(s) $\mathcal{L'}$, whose incommensurability with the underlying lattice $\mathcal{L}$ sets the quasiperiodicity of the bond occupations. In our setup, we choose $\mathcal{L}$ to be the square lattice whose bond lattice is $\mathbb{Z}^2$, and $\mathcal{L'}$ is a rotated square lattice with primitive lattice vectors $\vec{a}$ and $\vec{a^\perp}$ in the basis of $\mathbb{Z}^2$. The second decision is the definition of a continuously tunable parameter to vary the bond occupations, analogous to the probability $p$ of occupying a bond in random percolation models. In this section, we introduce the two different self-dual models used in this work, before elaborating on the "maximally incommensurate" choice of $\vec{a}$.

\begin{figure}[hbtp]
\subfloat[]{
\includegraphics[width=\linewidth]{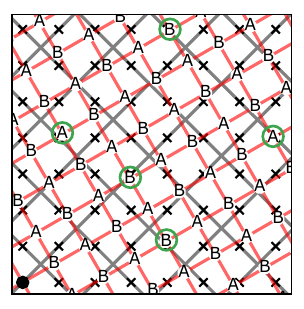}
\label{fig:checkerboard}
} \\
\subfloat[]{
\centering
\includegraphics[width=\linewidth]{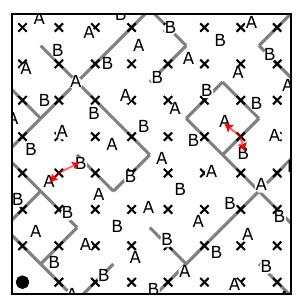}
 \label{fig:b0}
}
\caption{(a) Lattices $\mathcal{L}$ (gray), with bonds (black x's) positioned at the vertices of $\mathbb{Z}^2$, and $\mathcal{L'}$ (red). The solid black circle indicates the point where $\mathcal{L'}$ coincides with $\mathbb{Z}^2$. For the checkerboard percolation model, the vertices of $\mathcal{L'}$ are marked $A$ and $B$ in a checkerboard fashion. $\mathcal{L'}$ is chosen to be the Ford lattice, with primitive lattice vector defined in~\autoref{eq:ford}. Green circles indicate near-commensurate points, corresponding to the green stars in the left subplot of~\autoref{fig:ford}. (b) Percolation instance of the checkerboard model produced by (a) at $b=0$. Occupied edges, for which $d_A(x,y)>d_B(x,y)$, are shown in gray. Red arrows show the displacements to the nearest A and B points for two edges of $\mathcal{L}$. Even bonds of $\mathcal{L}$ have slope $+1$ in these figures, while odd bonds have slope $-1$.}
\end{figure}

\subsection{Lattice construction}
 For both models, label each bond of $\mathcal{L}$ by the coordinates of its midpoint, the integer pair $(x,y)$. In the "checkerboard model," $\mathcal{L'}$ is divided into A and B sublattices (\autoref{fig:checkerboard}), and $d_{A}(x,y)$, $d_{B}(x,y)$ denote the distance to the nearest A (B) vertex of $\mathcal{L’}$. Then a given realization of the percolation model is a graph with the edge set (see~\autoref{fig:b0}):
\begin{equation}\label{eq:model}
\mathcal{E} = \{(x,y): b(x,y) \equiv d_B(x,y) – d_A(x,y) \leq b\}
\end{equation}
The quantity $b(x,y)$ is referred to as the "b-score" of the edge $(x,y)$.

\begin{figure}[t]
    \centering
    \includegraphics[width=\linewidth]{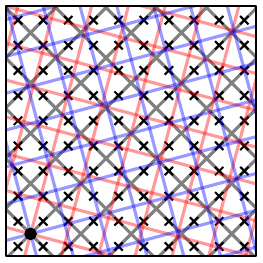}
    \caption{Lattices $\mathcal{L}$ (gray) with bonds (black x's) positions at the vertices of $\mathbb{Z}^2$, $\mathcal{L'}_A$ (blue), and $\mathcal{L'}_B$ (red), for the counter-rotated percolation model. $\mathcal{L'}_{A,B}$ are chosen to be counter-rotated Ford lattices (\autoref{eq:ford}), with primitive lattice vectors $\vec{a}_A=(\sqrt{3}/2, 1/2)$ and $\vec{a}_B=(\sqrt{3}/2,-1/2)$.}
    \label{fig:lattice-counter}
\end{figure}

The second model uses two lattices, $\mathcal{L'}_A$ and $\mathcal{L'}_B$, which are counter-rotated with respect to $\mathcal{L}$; i.e. $\vec{a}_A = (a_1, a_2)$ and $\vec{a}_B = (a_1, -a_2)$ (\autoref{fig:lattice-counter}). Taking $\mathcal{L}$, $\mathcal{L'}_A$, $\mathcal{L'}_B$ to all intersect at one point, a given realization of the model is again defined by~\autoref{eq:model}. We call this the "counter-rotated model."

Since A and B vertices are on even footing, both models are self-dual by design, with percolation transitions at $b_c=0$. Since all the lattices involved are square lattices, the models also have four-fold rotational symmetry. At the self-dual point, the counter-rotated model also has reflection symmetry through the horizontal, vertical, and $\theta=\pi/4$ axes, unlike the checkerboard model. The four-fold rotational symmetry can be broken by ranking the two bond orientations (slope $\pm 1$ in~\autoref{fig:checkerboard}) differently, thus producing a tunable two-fold anisotropy studied in~\autoref{sect:anisotropy}.

As defined, these models have no randomness: having fixed the lattices and the parameter $b$,~\autoref{eq:model} defines a deterministic rule for which bonds are present. An ensemble of percolation instances is formed by taking randomly chosen finite patches within a single, infinite quasiperiodic system. In the counter-rotated model, this random choice is manifested by taking the origin of each sample to be at some random, far displacement from the point where the three lattices to coincide. In the checkerboard model, setting the origin at a random distance from the intersection between $\mathcal{L}$ and $\mathcal{L'}$ is (morally) equivalent to translating $\mathcal{L'}$ by a random two-component vector with respect to $\mathcal{L}$.\footnote{Explicitly, the enlarged unit cell of the checkerboard has primitive lattice vectors $\vec{a} + \vec{a}^\perp$ and $\vec{a} - \vec{a}^\perp$. Then let $r_1, r_2$ be random numbers between 0 and 1, and displace $\mathcal{L'}$ by $\vec{R}=r_1 \vec{a} + r_2 \vec{a}^\perp$ to define one sample.}

Crucially, for a fixed translation of the lattice, if $(x,y) \in \mathcal{E}$ for some $b_0$, then it also belongs to the edge set for all $b>b_0$. This means that as we increase $b$, we only add bonds and never remove them. This helps avoid non-monotonicities in the scaling functions that arise in some other quasiperiodic models that we have examined. It also simplifies the algorithm for analyzing fixed $n$ ensembles, where $n$ is the fraction of bonds present. With $n$ as the tuning parameter, self-duality of the lattice implies $n_c=1/2$. For a given $\mathcal{L'}$, we perform a sweep in $n$ from 0 to 1 by adding bonds in order of increasing $b(x,y)$.

To probe universality, we also define a tunable family of counter-rotated models with the same symmetry by assigning each bond a "c-score",
\begin{equation}\label{eq:cscore}
    c(x,y) = b(x,y)(a - b(x,y)^2)
\end{equation}
where $a>0$ parameterizes members of the family. Percolation instances defined by occupying bonds with $c(x,y)\leq c$ have a self-dual point at $c=0$. For $a>\max(b(x,y)^2)=1/2$, $c(x,y)$ is a monotonic function of $b(x,y)$, so bonds are added in the same order as the standard counter-rotated model. However, for $a<1/2$, $c(x,y)$ is non-monotonic in $b$, so bonds are added in a different order. For sufficiently small $a$, this is found to modify the properties of the critical clusters at the percolation threshold.

\subsection{Maximally incommensurate lattices}
The above constructions produce a family of models, parameterized by the vector $\vec{a}$. To obtain models with smooth monotonic critical behavior, we want to avoid commensurate points between $\mathcal{L'}$ and $\mathbb{Z}^2$. One choice for $\mathcal{L'}$ that is in some sense "optimal" is:
\begin{equation}\label{eq:ford}
\vec{a}= (\sqrt{3}/2, 1/2)
\end{equation}
which we dub the "Ford lattice" in reference to the 1925 proof in which this lattice is used~\cite{Ford1925}. By symmetry, $\vec{a}=(\sqrt{3}/2, -1/2)$ is also an optimal choice, which defines $\mathcal{L'}_B$ in our counter-rotated construction. In this section, we define the sense in which these lattices is "maximally incommensurate" and present an algorithm for finding other optimal lattices.

Consider all the points of the superlattice $\mathcal{L}'$ within a distance $r/a$ of the origin, where one point in $\mathcal{L}'$ and one point in $\mathbb{Z}^d$ coincide with the origin.  In $d$ dimensions, there are $\sim (r/a)^d$ such points, each of which can be thought of as living within the unit cell of $\mathbb{Z}^d$. If these points are well spread out, the closest approach of any of these points (excluding the origin) to a point in $\mathbb{Z}^d$ is typically within distance $d \sim a/r$. This motivates us to use the quantity $d r/a$, where $d$ is the distance to the closest vertex of $\mathbb{Z}^d$, as our metric for how close to commensurate a point at distance $r$ is. Our goal is to construct $\mathcal{L'}$ such that $dr/a$ remains above some constant $C$ for all points other than the origin.

\begin{figure}[t]
\centering
\includegraphics[width=\linewidth]{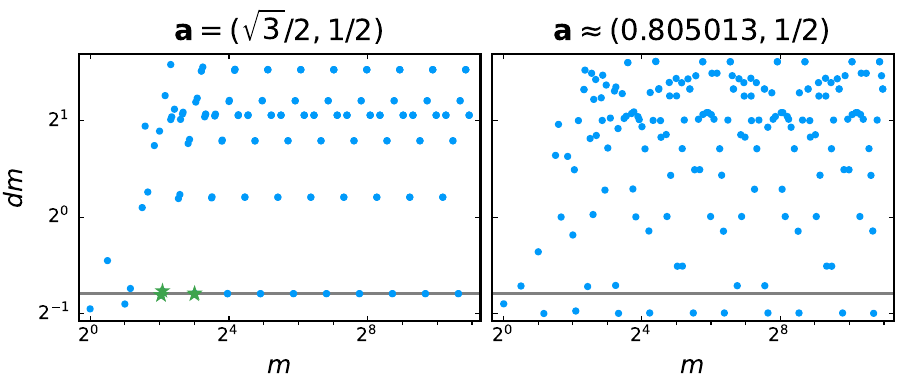}
\caption{Pattern of points with $dm<10$ on the Ford lattice (left) vs. one other choice of "optimized" parameters (right), following the protocol described in the text. The gray line indicates the Ford bound of $C=1/\sqrt{3}$. Green stars indicate the the near-commensurate points circled in~\autoref{fig:checkerboard}.}
 \label{fig:ford}
\end{figure}

In 1D, the best we can do is $C=1-1/\phi$, which is attained by choosing $a=\phi-1$, where $\phi$ is the golden ratio. In two dimensions, we seek $\vec{a}=(a_1,a_2)$ such that the lattice specified by primitive lattice vectors $\vec{a}$ and $\vec{a}^\perp = (-a_2, a_1)$ has the largest possible value for the closest approach, $\min[dr/a]$. More precisely, let $\vec{A}=\begin{pmatrix} 
a_1 & -a_2 \\
a_2 & a_1 
\end{pmatrix}$, so that a superlattice vertex with integer coordinates $\vec{m}=(m_1,m_2)$ has coordinates $\vec{A} \vec{m}$ on the underlying lattice. Letting $m\equiv \norm{\vec{m}}$, out to some maximum radius $m^*$, we compute
\begin{equation}
    dm_{\min}\equiv \min_{\vec{m}} d(\vec{m}) m
\end{equation}
Viewing $\vec{a}$ as a complex irrational number $\alpha$, this maps onto the problem of maximizing the quantity
\begin{equation}
k(\alpha) = \min_{p, q} |\alpha - p/q| |q|^2,
\end{equation}
where $p$ and $q$ are both complex numbers whose real and imaginary parts are integers. This in turn relates to a problem long answered in the mathematics literature~\cite{Ford1925,Perron1930,Schmidt1967}: for any $\alpha$, the inequality $k(\alpha) < C$ has infinitely many solutions $p, q$, if $C \geq 1/\sqrt{3}$. Thus, for any choice of $\vec{a}$, we will encounter infinitely many points for which $d(\vec{m})m \leq 1/\sqrt{3}$. On the other hand, Ref.~\cite{Ford1925} proves that if $C<1/\sqrt{3}$, then there exists a dense set of $\alpha$ for which the inequality only has a finite number of solutions ("near-commensurate points"). One such $\alpha$ is
\begin{equation}\label{eq:fordb}
\alpha = e^{i\pi/6} \Rightarrow \vec{a}= (\sqrt{3}/2, 1/2)
\end{equation}
which is our Ford lattice. Although a finite number of near-commensurate points would still be problematic for our purposes, the Ford lattice enjoys the property that as $m$ increases, the points closest to being commensurate saturate the Ford bound and occur in a scale-invariant pattern (left panel of~\autoref{fig:ford}). This makes the model well-behaved near the percolation transition, free of sharp non-monotonicities in the scaling behavior.

The distance where the Ford lattice comes closest to commensurate by this measure is therefore at small $m$, with $dm_{min} = dm(1,0) \approx 0.518$. To improve $\mathcal{L'}$ at small $m$, we search $(a_1,a_2)$ in the octant of the $\mathbb{Z}^2$ unit cell defined by $1-a_2 \leq a_1 \leq 1$, $a_2\leq 1/2$. Everything is modulo $\mathbb{Z}^2$, which has translation, rotation and reflection symmetry, so all other octants are equivalent by symmetries. We fix $a_2$ to lie on the boundary of this octant, $a_2=1/2$, and optimize $a_1$ according to the following strategy. First, we fix a threshold of $C=0.5$ and for each $\vec{m}$ in order of increasing $m$, exclude the interval of $a_1$ for which $dm<C$. Applied up to $m=100$, this procedure yields a few candidate values for $a_1$. A candidate $a_1$ is then further optimized by iterating the following steps:
\begin{enumerate}
\item Identify the smallest $m^*$ for which $dm^*<\min_{\vec{m}:m<m^*} dm$.
\item Either increase or decrease $a_1$ until $dm<dm^*$ for some $\vec{m}: m<m^*$, then fine-tune $a_1$ so that $dm=dm^*$.
\end{enumerate}
One version of this protocol, in which $a_1$ is always increased in step 2, yields lower convergents to $a_1=\sqrt{3}/2$, recovering the Ford lattice. More generally, each iteration of the algorithm yields a rational $a_1$ with a large denominator, so that the first commensurate point occurs at large $m$. One of these is $a_1=37805/46962\approx 0.805013$, whose pattern of close approaches is compared to Ford in~\autoref{fig:ford}. This other choice of $a$ does better at $m=1$, but its structure of near-commensurate points is less clean, failing to converge to the Ford bound at large $m$, and its $dm_{min}\cong 0.501$ is lower than that of the Ford lattice. {We have not systematically searched the interior of the octant of the unit cell, but every indication is that the largest $dm_{min}$ points occur on the boundary, where we have searched systematically.}

\subsection{Boundary conditions}
Using the scale-invariant pattern of near-commensurate points on the Ford lattice, we can also construct a series of rational approximants to $\vec{a}$ for use in periodic boundary conditions (PBCs). The near-commensurate points of the Ford lattice come in two sequences: an odd sequence of points near $[L_1^{o}, L_1^{o}]$, and an even sequence near $[L_1^e, 0]$. This motivates the use of two sequences of system sizes, for which we define the system on a bounding box with lattice vectors $[L_1,L_2]$ and $[-L_2,L_1]$ and perturb slightly away from the Ford parameters such that coordinates separated by integer multiples of these lattice vectors are equivalent.\footnote{While one might worry that even a slight perturbation could spoil the optimality, the chosen perturbation only creates a single commensurate point within the finite patch of interest; see~\appref{app:boundary}.} In both sequences, $L_1$ is defined recursively as:
\begin{equation}\label{eq:recursive-L}
    L_1(j) = 4 L_1(j-1) - L_1(j-2)
\end{equation}
The first sequence, labeled as the odd parity sequence, has $L_1=L_2$ and the initial conditions $L_1^{o}(1)=2, L_1^{o}(2)=8$. The opposite parity sequence has $L_2=0$ with the initial conditions $L_1^{e}(1)=6, L_1^{e}(2)=22$~\cite{oeis}.

Further details on open and periodic boundary conditions are provided in~\appref{app:boundary}. Where possible, we use the methods described in the next section to work on a "nominally infinite" lattice, thus avoiding the need to engineer commensurate points. 

\section{Methods}\label{sect:methods}
In this section we describe the numerical methods for analyzing our quasiperiodic models.

\subsection{Nominally infinite methods}
A striking property of both models is the existence of a discrete sequence of fractal clusters at the percolation threshold, rather than the emergent conformal invariance characteristic of the random percolation critical point. To identify as many generations of these discrete sequences as possible, we adapt methods for growing a single cluster or its external hull on a nominally infinite lattice. These methods are discussed in turn.

\subsubsection{Incremental cluster growth}
The method for building clusters up to $b_c=0$ on nominally infinite lattices is an adaptation of the Leath method~\cite{Leath1976}, which views percolation as a "wetting"~\cite{Vyssotsky1961} or "epidemic"~\cite{Grassberger1983} process by growing an individual cluster from a single seed site. The standard algorithm is conducted at a fixed probability $p$, and in terms of the b-scores defined in the previous section can be phrased for bond percolation as follows~\cite{Ziff2021}. A "pocket" $\mathcal{P}$ is initialized with a single "wet" vertex $\vec{v}_0$, and the cluster $\mathcal{C}$ is initially empty. Iteratively until the pocket is empty, a vertex $\vec{v}$ is popped from $\mathcal{P}$ and added to $\mathcal{C}$. For each incident edge that has not already been visited, we assign a b-score $b(x,y)$ (a random number between 0 and 1). If $b(x,y)>p$, the bond is made unoccupied. Otherwise, the bond is occupied, so its other endpoint $\vec{u}$ becomes wet and is added to the pocket if it has not been already. The algorithm therefore terminates when all the edges between wet and dry vertices have been visited and made unoccupied. The same algorithm could be used for nonrandom percolation, with $b(x,y)$ assigned according to a given deterministic prescription.

\begin{algorithm}
\SetAlgoLined
\SetKwInOut{Output}{Output}
\SetKwInOut{Input}{Input}
\Input{Seed vertex $\vec{v_0}$ \\
Maximum b-score $b_0$
}
\Output{Cluster $\mathcal{C}$ containing $\vec{v_0}$ at $b=b_0$, with each site labeled by the b-score at which it joins the cluster}
$b \gets -\infty$\;
$\mathcal C \gets \{(\vec{v}_0,b)\}$\;
$\mathcal{B}\gets$ empty priority queue of "blocking edges"\;
\For{$(x,y)$ \normalfont{edge incident to} $\vec{v}_0$} {
Add $((x,y), b(x,y))$ to $\mathcal{B}$\;
}

\While{$b<b_0$}{
$\mathcal{B}, \mathcal{C}, b \gets$ grow($\mathcal{B}, \mathcal{C}, b_0$);
}
\Return $\mathcal{C}$;
\caption{Nominally infinite cluster growth\label{alg:infinite-clusters}}
\end{algorithm}

\begin{algorithm}
\SetAlgoLined
\SetKwInOut{Output}{Output}
\SetKwInOut{Input}{Input}
\Input{Priority queue of blocking edges $\mathcal{B}$ at step $i$ \\
Cluster $\mathcal{C}$ at step $i$ \\
Maximum b-score $b_0$
}
\Output{$\mathcal{B}$ at step $i+1$ \\
Cluster $\mathcal{C}$ at step $i+1$ \\
Current b-score $b$}
Edge $\vec{e}$, b-score $b \gets$ pop $\mathcal{B}$\;
\lIf{$b>b_0$}{\Return $\mathcal{B}, \mathcal{C}, b$}
$\mathcal{P} \gets$ \{endpoints of $\vec{e}$  not already in $\mathcal{C}$\}\;\label{line:pocket}
\While{$\mathcal{P} \neq \emptyset$} {
Vertex $\vec{v} \gets$ pop $\mathcal{P}$\;
Add $(\vec{v}, b)$ to $\mathcal{C}$\;
\ForEach{\normalfont{neighbor $\vec{u}$ of $\vec{v}$ not already in $\mathcal{P}$ or $\mathcal{C}$}} {
$(x,y) \gets $ coordinates of edge linking $\vec{u}$ and $\vec{v}$\;
\eIf{$b(x,y)\leq b$}{
Add $\vec{u}$ to $\mathcal{P}$\;}
{Add $((x,y), b(x,y))$ to $\mathcal{B}$\;\label{line:block}}
}}
\Return $\mathcal{B}, \mathcal{C}, b$\;
\caption{Incremental growth of cluster \texttt{grow}\label{alg:grow}}
\end{algorithm}

Taking cues from the Newman-Ziff algorithm discussed below~\cite{Newman2000,Newman2001} for performing efficient sweeps in $p$, we modify this algorithm to build clusters in the entire range of b-scores up to some $b_0$. The algorithm is given as pseudocode in~\autoref{alg:infinite-clusters} and the subroutine~\autoref{alg:grow}. Starting from $b=-\infty$, each time a bond is assigned a score greater than $b$, instead of just being set to "unoccupied," it is added to a priority queue of "blocking edges" (\autoref{line:block} of~\autoref{alg:grow}). The growth at a given $b$ ends when all bonds between wet and dry vertices are blocking edges. In each successive step, the blocking edge with the smallest b-score is removed and a pocket is initialized with its adjacent dry vertices (\autoref{line:pocket}). A full run of the algorithm completes when all blocking edges have b-scores $>b_0$. Taking $b_0=b_c=0$ therefore generates the fractal clusters at the critical point, whose mass $M$ scales as $M\propto r^{D_f}$ where $r$ is the cluster radius and $D_f$ the fractal dimension~\cite{Christensen2002}.

Recording the value of $b$ at which each vertex is made wet also gives a complete record of the clusters containing the seed vertex for the entire range of $b<b_c$. In practice, the algorithm may terminate before reaching $b_c$ if the cluster exceeds a prespecified mass threshold (chosen to avoid arbitrarily long runtime). However, since this algorithm only requires the calculation of b-scores for edges that may extend the single cluster, we can access much larger cluster sizes than with finite system methods.

\subsubsection{Identification of hulls}
While critical clusters are characterized by the fractal dimension $D_f$, their external perimeters ("hulls") are also fractal objects, with fractal dimension $D_h$. The external hull of an already identified cluster can be obtained using the method in Ref.~\cite{Voss1984}. In square lattice bond percolation, every point of $\mathbb{Z}^2$ is the midpoint of either an occupied bond of $\mathcal{L}$, or an occupied bond of its dual. A self-avoiding walk is traversed by hopping between nearest neighbors on $\mathbb{Z}^2$, turning clockwise (counterclockwise) if the bond belongs to the original (dual) lattice. To grow the hull "from scratch" in a similar spirit to the Leath method, without needing to find a full cluster first, the occupied/unoccupied status of a bond is determined on the fly~\cite{Ziff1984,Weinrib1985,Grassberger1986}. The walk terminates, and the hull closes on itself, when the origin of the walk is re-encountered facing in the same direction as the first step.

In traversing the hull we keep track of three related quantities. The "hull itself" is defined to consist of the midpoints of the links of the self-avoiding walk. This walk is the internal perimeter of a cluster on one lattice and the external perimeter of a cluster on its dual. Thus we can also associate to each hull a series of "inner" and "outer" bonds adjacent to it. At the percolation threshold, $N_{inner}, N_{outer}, $ and $N_{hull}$ all scale with a fractal dimension of $D_h$, i.e. $N \propto r^{D_h}$.

It should be noted that unlike the clusters they enclose, the external perimeters have the feature that as $b$ increases, some links are added to the hull but others are removed, as bonds change affiliation from "unoccupied" to "occupied". This prevents us from incrementally growing the hulls in an efficient manner, so we instead only obtain the hulls precisely at $b=b_c$. 

\subsection{Incremental percolation on finite systems}\label{sect:incremental}
Complementing these methods on nominally infinite lattices, we also collect data on finite lattices with periodic or open boundary conditions via the Newman-Ziff algorithm. For a given finite lattice $\mathcal{L}$ a percolation instance is constructed using a weighted union-find algorithm with path compression~\cite{Newman2000,Newman2001}. Each vertex $v$ of $\mathcal{L}$ is labeled with a parent $p$ within the same cluster and a displacement $\vec{d}(p\rightarrow v)$ from itself to its parent. A vertex with no parent is the root of the cluster, and keeps a record of the number of vertices belonging to the cluster. Thus, to determine whether two vertices belong to the same cluster, we simply traverse their respective trees of parents to find the roots. To make the algorithm more efficient, whenever the "find root" operation is performed, every vertex along the path to the root is relabeled to point directly to the root ("path compression").

The system is initialized with each vertex as the root of its own isolated cluster. The available edges are sorted according to their b-scores, then added incrementally in order of increasing $b$. For every added edge, connecting vertices $u$ and $v$, we find the roots $r(u)$ and $r(v)$ of their respective clusters. If $r(u)\neq r(v)$, we join the two clusters by making the root of the larger cluster the parent of the other root ("weighted union") and setting the mass of this merged cluster to be the sum of the individual cluster sizes. If $r(u)=r(v)$, this newly added edge does not merge any clusters, but it could complete a path that completely wraps around the boundaries. 

A given cluster can wrap around the system along the "$+$" direction (defined by the vector $[L_1, L_2]$) and/or along the "$-$" direction (defined by the vector $[-L_2, L_1]$). Wrapping clusters are identified using the algorithm developed in Ref.~\cite{Machta1996}, by computing the displacement along the loop from $u$ to $v$ along the new bond, $\vec{e}_{uv}=(\pm 1, 1)$, then back from $v$ to $r(v)$ to $u$. The total displacement is:
\begin{equation}
    \vec{d}(u\rightarrow v \rightarrow u) = \vec{e}_{uv} + \vec{d}(v\rightarrow r(v)) + \vec{d}(r(v)\rightarrow u)
\end{equation}
If the total displacement is not zero, then a wrapping event has occurred, and $\vec{d}(u\rightarrow v\rightarrow u)$ points along a lattice vector associated with the periodic boundary conditions. 

Since bonds are added in an incremental fashion, the precise bond rank (in terms of $n$ or $b$) of the first wrapping event along each direction can be identified. We can then define $\Pi^+(x)$ and $\Pi^-(x)$ as the probabilities that the sample contains a wrapping cluster in the $+$ or $-$ direction respectively (which are equivalent in the absence of anisotropy). The probability that two vertices belong to the same cluster decays exponentially in the non-percolating phase, with a correlation length $\xi$ that diverges as $(n-n_c)^{-\nu}$~\cite{Christensen2002}. Thus, in principle, the wrapping probability can be scaled according to
\begin{align}\label{eq:scaling-wrap}
    \Pi_{wrap}(n, L) &= f(L^{1/\nu} (n-n_c)) \notag \\
    \Pi_{wrap}(b, L) &= g(L^{1/\nu} (b-b_c))
\end{align}
for scaling functions $f(x), g(x)$.

Note that wrapping events can never occur with open boundary conditions; instead, we look for "spanning clusters" which contain vertices belonging to opposite boundaries. The finite-size corrections to scaling are generally more severe for the spanning clusters than for the wrapping probability, so $\Pi_{span}$ is inferior to $\Pi_{wrap}$ for determining $\nu$~\cite{Skvor2007,Mertens2012}. However, since the available system sizes and orientations for PBCs are constrained by the commensurability of $\mathcal{L}$ and $\mathcal{L'}$, the crossing probabilities provide a useful alternative observable, particularly when we add two-fold anisotropy in~\autoref{sect:anisotropy}.

\section{Fractal clusters}\label{sect:fractal}

By constructing instances of the Ford lattice percolation models in finite and nominally infinite systems, we now determine the exponents $D_f$ and $D_h$ which describe the critical clusters at $n_c = 1/2$. We find stark qualitative and quantitative differences from the cluster size distribution for random percolation, in both models.

\subsection{Checkerboard model}
First we consider the distribution of cluster sizes for the checkerboard model, on finite system sizes with periodic boundary conditions. 
Despite being produced by different incommensurate translations of $\mathcal{L'}$ with respect to $\mathcal{L}$, samples fall into just a few groups at $b=0$, whose connected components are identical up to a translation by a lattice vector of $\mathcal{L}$.

\begin{figure}[t]
    \subfloat[]{
    \includegraphics[width=0.48\linewidth]{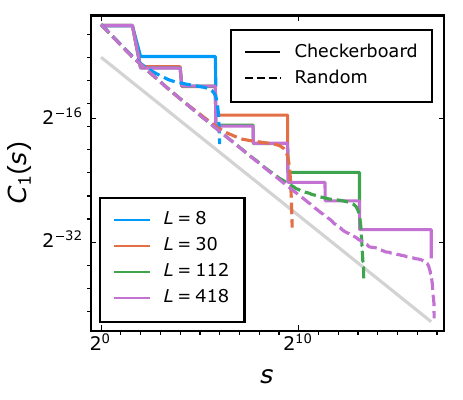}\label{fig:cumul}
    }
\subfloat[]{
    \includegraphics[height=0.45\linewidth,width=0.45\linewidth,keepaspectratio]{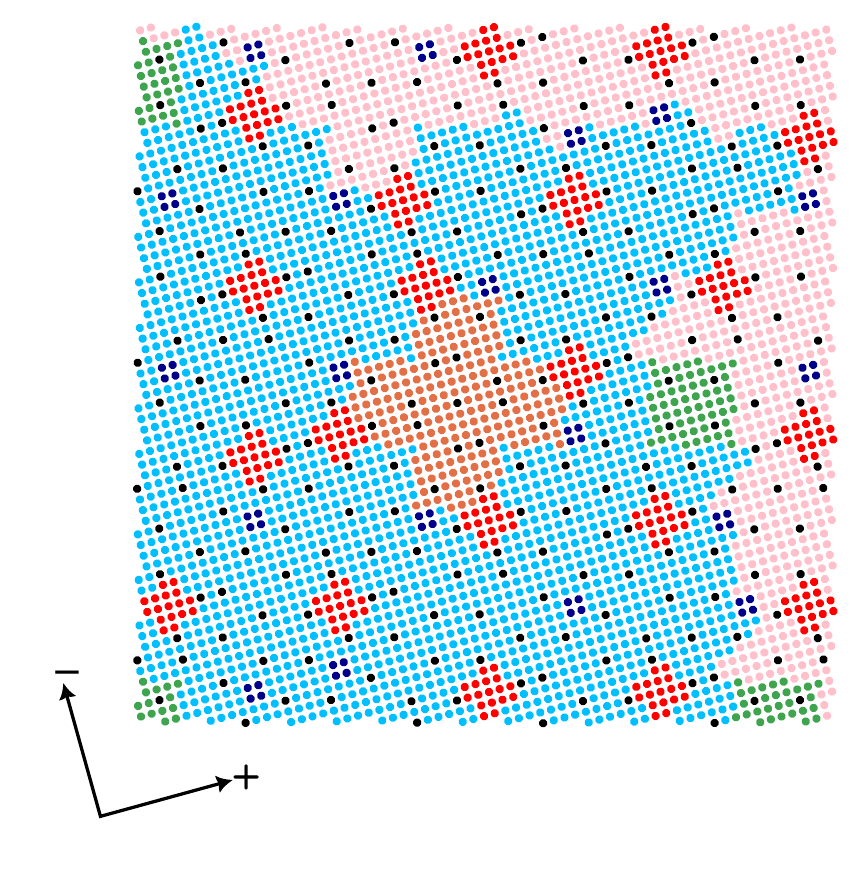}
    \label{fig:fractal}
    }
\caption{(a) Cumulative critical distribution of cluster sizes $s$ sites in the checkerboard Ford model, for population I of $[L_1^o, L_1^o]$ system sizes (wrapping threshold $b_c>0$). Dashed lines: critical random percolation, with thin gray line showing power-law decay with $\tau=187/91$. (b)  Zoom-in on one quadrant of a critical ($n_+=n_-=1/2$) percolation cluster (blue) of mass $8632$ sites exhibiting four-fold symmetry, showcasing the fractal structure of smaller four-fold symmetric connected clusters (different colors) that are also present. Axes indicate the orientations of $+/-$ bonds on the underlying lattice, relative to the cluster.  {See~\appref{app:fractals} for more illustrations of the  critical clusters.}\label{fig:clusters}}
\end{figure}

Of the three populations, defined explicitly in~\autoref{sect:nu}, population I of samples with system sizes of the form $[L_1^o, L_1^o]$ consists of samples where the wrapping threshold is strictly positive, i.e. the clusters at $b=0$ do not wrap around either boundary. In~\autoref{fig:cumul}, we plot the cumulative cluster size distribution for this population, defined as:
\begin{equation}
C_1(s) = \sum_{s'=s}^{s_{max}} n(s')/s'
\end{equation}
where $n(s')$ is the number density per lattice site of clusters of mass $s'$ sites, and $s_{max}$ is the mass of the largest cluster. This distribution is overlaid with that of the random percolation model, which has a power law tail of slope $-\tau=-1-d/D_f$ and a rounded "shoulder" at $s\sim L^{D_f}$~\cite{Stauffer1979}. In contrast, the checkerboard Ford distribution consists of a series of finite steps; only after smoothing this "staircase" would we have a power-law tail similar to the random case.

The clusters comprising these steps belong to two species, all possessing four-fold symmetry. As illustrated in~\autoref{fig:fractal}, successive generations of each species contain holes occupied by earlier generations. More details on this fractal structure are provided in~\appref{app:fractals}; we have not been able to discover the analytic form for the two sequences of cluster sizes, so we leave this as an unsolved puzzle to challenge the interested reader.

Random percolation, for example on a square lattice, has an emergent conformal symmetry at its critical point, where the large connected clusters have a probability distribution that asymptotically has continuous scale invariance and full rotational symmetry (which are familiar subgroups of conformal symmetry)~\cite{Langlands1994,Smirnov2001,Duminil-Copin2020}. Our quasiperiodic model's critical point, on the other hand, is manifestly {\it not} conformally invariant, since the large connected clusters only have a discrete scale invariance and do not have any emergent continuous rotational invariance; they only have the four-fold rotational invariance that is there already microscopically.

Having identified these fractal clusters, their fractal dimension can be determined to excellent precision and is found to be higher than that of random percolation. For this we use the nominally infinite lattice methods. From the definition $M(r) \propto r^{D_f}$ for a cluster of mass $M$ and linear dimension $r$, the inferred exponent from two successive generations of a given sequence is
\begin{equation}\label{eq:df-j}
D_f(j, j-1) = \frac{\log[M(j)/M(j-1)]}{\log[r(j)/r(j-1)]}
\end{equation}
This asymptotes towards the "true" fractal dimension in the limit of $j\rightarrow \infty$. 

\begin{figure}[t]
    \centering
    \subfloat[]{
    \includegraphics[width=0.46\linewidth]{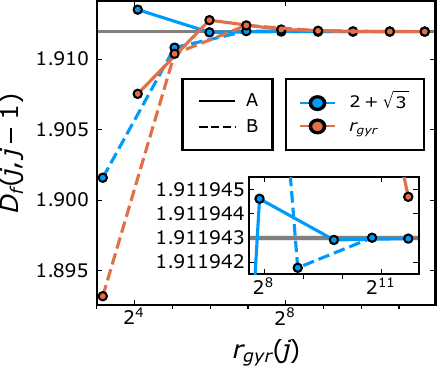}
    \label{fig:df-checker}}
    \subfloat[]{
        \includegraphics[width=0.46\linewidth]{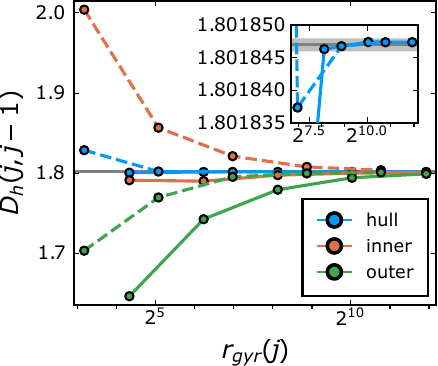}
    \label{fig:dh-checker}}
    \caption{(a) Fractal dimension $D_f$ and (b) fractal hull dimension $D_h$ inferred from the scaling of A (solid lines) and B (dashed lines) clusters/hulls in the checkerboard Ford model. In (a), blue curves come from using $r(j)/r(j-1)=2+\sqrt{3}$ in~\autoref{eq:df-j}, while orange curves use $r(j)=r_{gyr}(j)$. In (b), blue, orange, and green curves use the number of hull links, inner bonds, and outer bounds respectively as $N(j)$ in~\autoref{eq:dh-j}, with $r(j)/r(j-1)=2+\sqrt{3}$ for all curves.\label{fig:fractal-cluster}}
\end{figure}

In~\autoref{eq:df-j} we need to define a measure of the mass $M$ and radius $r$ of each cluster. For the mass we use the number of sites (which is empirically observed to lead to faster convergence than using the number of edges). For the radius, a standard choice is the radius of gyration:
\begin{equation}
    r_{gyr}(j) = \sqrt{I(j)/M(j)}
\end{equation}
where $I(j)$ is the moment of inertia of the $j$th generation cluster. The values of $D_f$ inferred from the scaling of mass with radius of gyration are plotted in orange in~\autoref{fig:df-checker} for the A and B sequences. Also shown in~\autoref{fig:df-checker} is a series of estimates $D_f(j,j-1)$ which converges slightly faster, obtained by substituting the asymptotic scale factor between successive generations:
\begin{equation}\label{eq:scale}
    \lim_{j\rightarrow \infty} \frac{r(j)}{r(j-1)} = 2+\sqrt{3}
\end{equation}
This scale factor originates from taking $\mathcal{L'}$ to be the Ford lattice, whose near-commensurate points are discretely scale-invariant with the same scale factor. Our resulting estimate of $D_f$ is:
\begin{equation}\label{eq:df-checker}
    D_f = 1.911943 \pm 10^{-7}
\end{equation}
in contrast to the random percolation value of $91/48=1.89583...$.

A similar method is used to determine the fractal dimension of the hulls, examples of which are shown in~\autoref{fig:checker-hull} of~\appref{app:fractals}. %A given hull can be defined as a closed path (shown in blue in~\autoref{fig:checker-hull}) in between the occupied bonds on one lattice (shown in black) and the occupied bonds on its dual lattice (shown in red). 
A given hull can be defined as a closed path composed of links in between the occupied bonds on one lattice and the occupied bonds on its dual lattice. Then the number of inner bonds, outer bonds, and hull links all scale as $N(j) \propto r(j)^{D_h}$. Thus the inferred fractal hull dimension is: 
\begin{equation}\label{eq:dh-j}
    D_h(j, j-1)=\frac{\log[N(j)/N(j-1)]}{\log[r(j)/r(j-1)]}
\end{equation}
As with the fractal cluster dimension, we obtain the fastest convergence by substituting $2+\sqrt{3}$ for $r(j)/r(j-1)$. A conservative estimate of $D_h$ is:
\begin{equation}
    D_h = 1.801847 \pm 10^{-6}
\end{equation}
significantly larger than the random percolation value of $D_h=7/4$.

\subsection{Counter-rotated model}
\begin{figure}[hbtp]
    \centering
    \subfloat[]{
    \includegraphics[width=\linewidth]{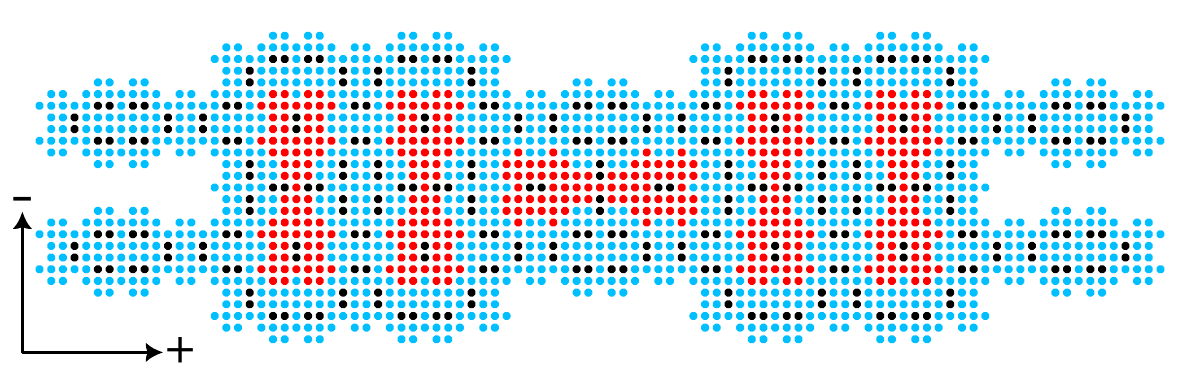}
    \label{fig:counter-cluster}} \\
    \subfloat[]{
    \includegraphics[width=\linewidth]{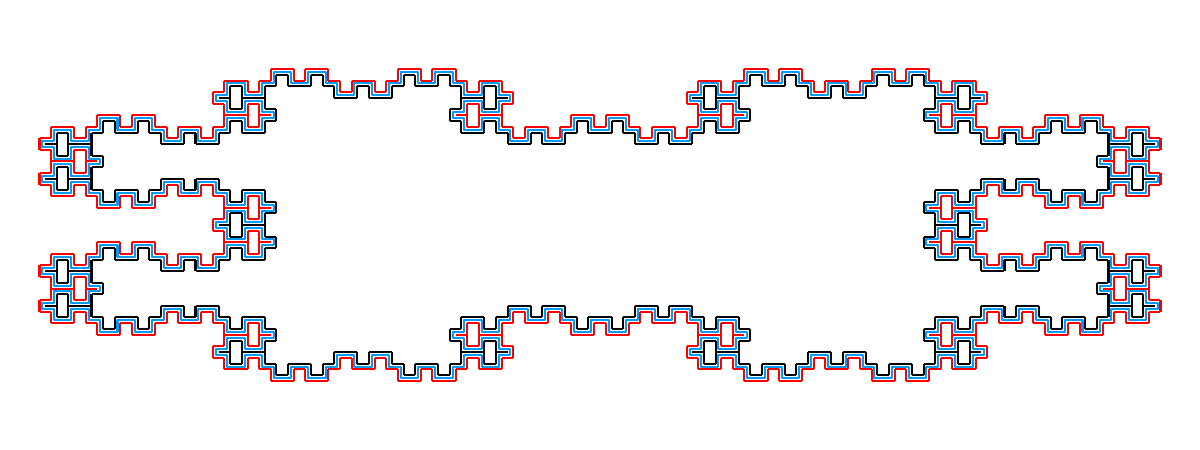}
    \label{fig:counter-hull}}
    \caption{(a) Fractal cluster of mass 1244 and (b) its external hull in the counter-rotated model. In both panels, the bonds of $\mathcal{L}$ have been aligned with the horizontal and vertical. In (b), black, red, and blue lines respectively indicate the inner bonds, outer bonds, and "hull itself" as defined in the text. }
    \label{fig:counter}
\end{figure}

Turning to the counter-rotated model, we again obtain a discrete sequence of fractal clusters at the critical point, with the scale factor $2+\sqrt{3}$. Unlike the checkerboard model, there is only one species of cluster, the fourth generation of which is shown in~\autoref{fig:counter-cluster}. Since the individual clusters have only a two-fold rotational symmetry, each generation in the sequence consists of two clusters which are mirror images of each other. Thus, the total population of critical clusters again only has the four-fold rotational symmetry that is microscopically present. The population of critical clusters is also invariant under reflections about the horizontal and vertical axes, axes of symmetry present at the critical point of the counter-rotated model but not the checkerboard model.

Compared to the checkerboard model's critical clusters, these "belt-buckle" clusters are significantly less dense than the "pinwheel" clusters of the checkerboard model, and the contrast between the hulls of the respective models is even starker (\autoref{fig:counter-hull}). Using \autoref{eq:df-j} and \autoref{eq:dh-j}, we arrive at the following estimates for $D_f$ and $D_h$:
\begin{align}
    D_f &= 1.707234 \pm 4 \cdot 10^{-6} \label{eq:df-counter}\\
    D_h &= 1.33850 \pm 5 \cdot 10^{-6} \label{eq:dh-counter}
\end{align}

\begin{figure}[t]
    \centering
    \subfloat[]{
    \includegraphics[width=0.46\linewidth]{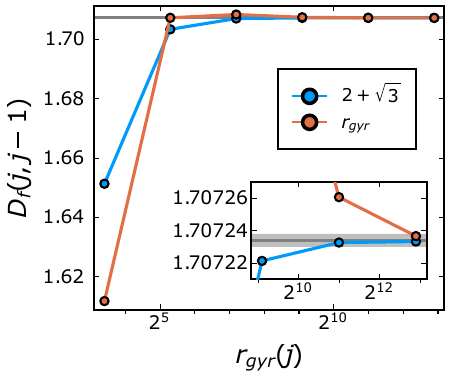}
    }
    \subfloat[]{
    \includegraphics[width=0.46\linewidth]{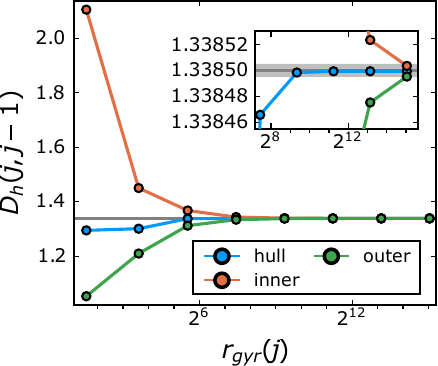}
    }
    \caption{(a) fractal dimension $D_f$ and (b) hull exponent $D_h$ in the counter-rotated Ford model. The colors are the same as in~\autoref{fig:fractal-checker}.}.
    \label{fig:fractal-checker}
\end{figure}

Through the formulation of random percolation as a $Q\rightarrow 1$ Potts model, the hull fractal dimension $D_h$ is known to satisfy~\cite{Saleur1987}:
\begin{equation}\label{eq:hull-nu}
    D_h = 1 + 1/\nu
\end{equation}
As we will see in~\autoref{sect:nu}, neither of our quasiperiodic models obeys this relation. The counter-rotated model is a particularly clear example of this, as~\autoref{eq:hull-nu} would imply $\nu>\nu_{random}$ whereas we instead obtain clear evidence of the opposite. This should come as no surprise, since in the absence of quenched randomness, the mapping to a Potts model no longer applies.

\subsection{c-scores and lack of universality}
\begin{figure}
    \centering
    \includegraphics[width=\linewidth]{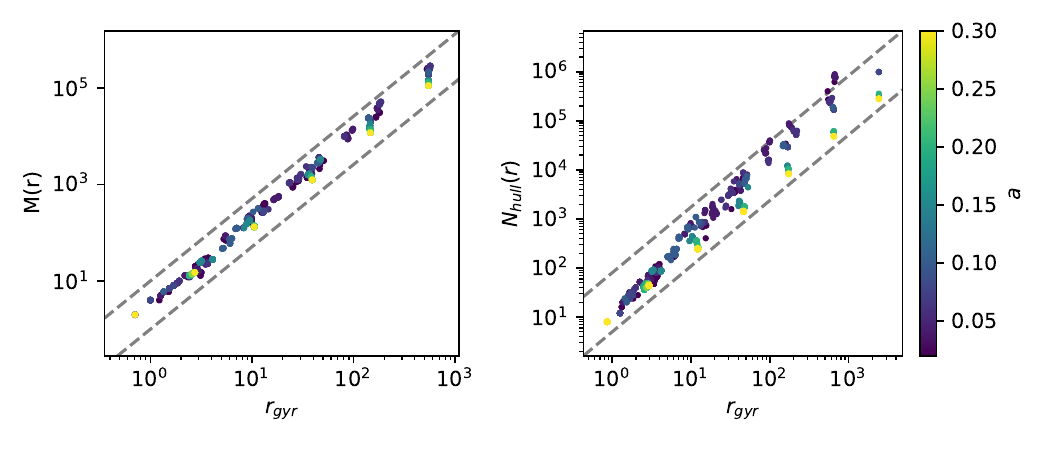}
    \caption{Scaling of cluster mass (left) and hull length (right) vs. radius of gyration for variations of the counter-rotated model tuned by $a$ (\autoref{eq:cscore}). In both panels, 200 samples were taken at each $a$ with an upper cutoff of $M=350000$ and $N_{hull}=10^6$, discarding samples that reached the cutoff. Dashed lines are power laws with $D_f=1.707243$ and $D_h=1.33850$.}
    \label{fig:cscores}
\end{figure}

In~\autoref{fig:cscores}, the critical clusters and hulls are plotted for the "c-score" variations on the counter-rotated Ford model (\autoref{eq:cscore}), with $a$ ranging from $0.02$ to $0.3$. For $a=0.3$, the clusters at the percolation threshold belong to the same discrete sequence as for the standard ($a \gg 1$) model. But as $a$ is decreased, two changes occur: the discrete sequence breaks up into a broader distribution of cluster sizes, and the fractal dimension changes. The latter trend is particularly noticeable for the hulls. Over the range of $a$ considered, $D_h$ tends to increase with decreasing $a$.\footnote{This only holds down to small but finite $a$, since for $a=0$ the original model is recovered, just with a rescaling $b\rightarrow -b^3$.} 

The sensitivity of $D_h$, and to a lesser extent $D_f$, to $a$ is evidence against universality. In the usual concept of universality classes of phase transitions, the universality class is the full set of exponents. The fractal dimensions of the critical clusters provide a good test of universality because the precision is best and the variations are large enough to cleanly see. As discussed in the next section, our estimate of $\nu$ is much less precise, and we focus solely on the original "b-score" models.

\section{Determination of $\nu$}\label{sect:nu}
Another key difference between the quasiperiodic and random percolation models is in the exponent $\nu$. In this section, we employ two different methods---one using the sequence of critical clusters on nominally infinite lattices, the other using an observable defined for periodic boundary conditions---to estimate $\nu$. Although the uncertainties are large, and the two methods yield somewhat different estimates for the checkerboard model, all estimates of $\nu$ fall significantly below its random percolation value of $4/3$.

\subsection{Scaling collapse on nominally infinite systems}
\begin{figure}[t]
    \centering
    \includegraphics[width=\linewidth]{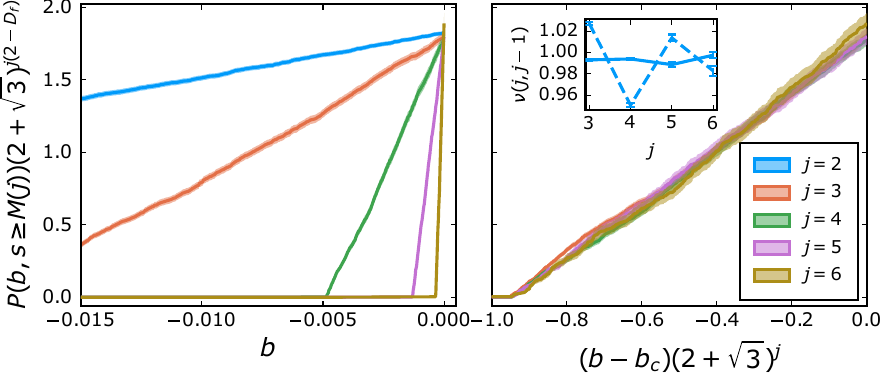}
    \caption{Scaling collapse for the counter-rotated model. Left: unscaled data, from generations 2-7 of cluster masses (\autoref{table:counter}). Right: scaling collapse with $\nu=1$, excluding $M(2)=17$. Value of $\nu$ inferred from scaling the slope of $y(b)$ between $0.01$ and $1.7$ (solid) and $[0.5,1.7]$ (dashed) (\autoref{eq:slope-nu}) is shown as an inset.}
    \label{fig:nu-counter}
\end{figure}

Our method for determining $\nu$ from the incremental cluster growth on nominally infinite lattices relies on the following scaling hypothesis for the probability of hitting on a cluster of mass $\geq S$~\cite{Ziff2021}:
\begin{align}\label{eq:scaling}
P(s\geq S, p) &= c_0 S^{2-\tau} f(c_1 (p-p_c) S^\sigma) \notag \\
&= c_0 S^{1-d/D_f} f(c_1(p-p_c) S^{1/\nu D_f})
\end{align}
By definition $\sigma=1/\nu D_f$ and the scaling relation $\tau = d/D_f + 1$ still holds for deterministic percolation (where our tuning parameter is $b$ or $n$ instead of $p$)~\cite{Stauffer1979}. \comment{Indeed, if instead of asking about the mass S we look at the probability of hitting a cluster of length $\geq L$, then since $S \propto L^{d_f}$ we get
\begin{equation}
P(l \geq L, p) \propto L^{D_f-d} g(x)
\end{equation}
where the scaling function g has the familiar argument of $x= (p-p_c) L^{1/\nu}$.}
\autoref{eq:scaling} was used to determine $p_c$ for random percolation in 3D models~\cite{Lorenz1998,Lorenz2000}, where $\sigma$ and $\tau$ had already been determined. Conversely, in our deterministic models, we know $b_c=0$, $n_c=1/2$, and have a very good determination of $D_f$, so a scaling collapse of $P(s \geq S, b) S^{2/D_f-1}$ vs. $b S^{1/\nu d_f}$ for different $S$ yields an estimate of $\nu$. 

The scaling collapse for the counter-rotated model is shown in~\autoref{fig:nu-counter}. Choosing $S$ to be the masses $M(j)$ of the fractal clusters at $b=0$, and using the fact that $M(j) \propto (2+\sqrt{3})^j$,~\autoref{eq:scaling} can be rewritten as:
\begin{align}\label{eq:scaling-b}
    y(b,j) &\equiv P(s \geq M(j), b) (2+\sqrt{3})^{j(2-D_f)} \notag \\
    &= g(b (2+\sqrt{3})^{j/\nu})
\end{align}
The scaling function $g(x)$ is roughly linear near the critical point, so $\nu$ can be estimated from scaling the slope $m$ of the unscaled data $y(b,j)$ vs. $j$:

\begin{equation}\label{eq:slope-nu}
    \nu(j,j-1)=\frac{\log[2+\sqrt{3}]}{\log[m(j)/m(j-1)]}
\end{equation}
This yields a critical exponent of
\begin{equation}
    \nu = 1.0 \pm 0.05
\end{equation}
The right panel demonstrates a good scaling collapse with $\nu=1$. The inset shows the inferred $\nu(j,j-1)$ from the slope in the interval $y\in[0.01, 1.7]$ (solid) and $y \in [0.5,1.7]$ (dashed). The inferred exponent is somewhat sensitive to the range of the fit, but remains within the stated uncertainty of $\pm 0.05$. A consistent estimate of $\nu$ is obtained taking $n$ as the tuning parameter rather than $b$, since $n(b)$ is a smooth function in the infinite system size limit.\footnote{See~\autoref{eq:nb} in~\appref{app:fss} for an explicit expression for $n(b)$.}

\begin{figure}[t]
    \centering
    \includegraphics[width=\linewidth]{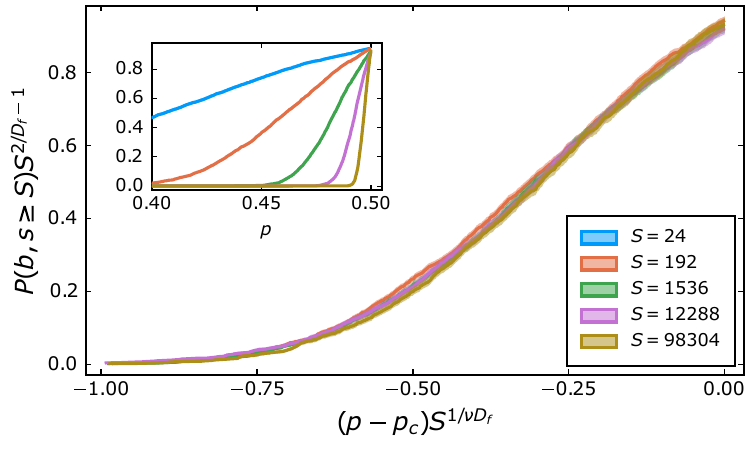}
    \caption{Scaling collapse for random percolation according to~\autoref{eq:scaling}, with $D_f = 91/48,\nu=4/3$. Inset shows unscaled data, including $S=24$ which is omitted from the scaling collapse.}
    \label{fig:nu-random}
\end{figure}

To confirm the validity of this method, the data for random percolation are shown in~\autoref{fig:nu-random}, using a sequence of masses with scale factor of 8. An excellent scaling collapse is obtained with $\nu=4/3$ and $D_f=91/48$, the known exponents for random 2D percolation.

\begin{figure}
    \centering
    \includegraphics[width=\linewidth]{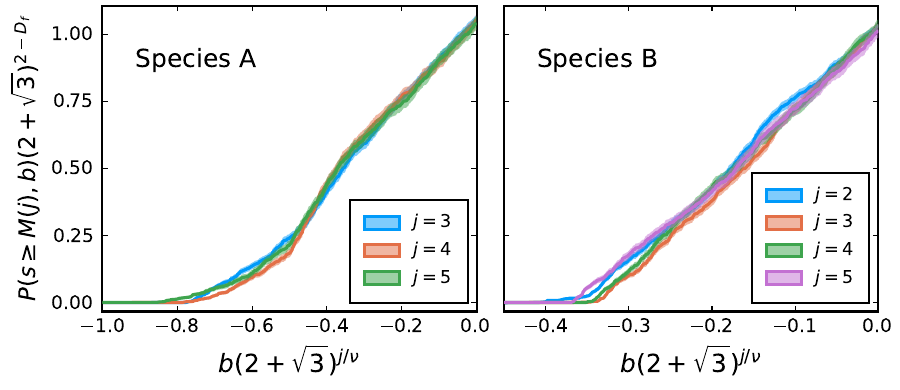}
    \caption{Scaling collapse of~\autoref{eq:scaling} with $\nu=0.95$ for species A (left, cf.~\autoref{table:a}) and B (right, cf.~\autoref{table:b}) of critical clusters in the checkerboard Ford model.}
    \label{fig:nu-checker}
\end{figure}

For the checkerboard model, we perform a separate scaling collapse for the A and B sequences of cluster sizes (\autoref{fig:nu-checker}). The best overall scaling collapse is obtained with $\nu\approx 0.95$, but $\nu(j,j-1)$ is not converging very well with $j$. In particular, while the curves for $M_B(3)=208$ and $M_B(4)=2576$ of sequence B coincide nicely, the $M_B(5)=31952$ curve (purple) is less steep, suggesting a larger $\nu$. Thus, in the next subsection we study the behavior on finite systems to obtain an alternative estimate of $\nu$.

\subsection{Distribution of wrapping events}
%Having seen a lack of universality in the previous section, for largely autobiographical reasons in the remainder of the paper we focus on the checkerboard model.
Standard methods for determining $\nu$ come from finite size scaling of observables on tori, such as the wrapping probability (\autoref{eq:scaling-wrap}). With the incremental method described in~\autoref{sect:incremental}, we can identify the number of occupied bonds at each "wrapping event" in individual samples. The first wrapping event occurs when a cluster wraps in one of the directions of periodic boundary conditions. Zero or more bonds are then added before this percolating clusters also wraps in the orthogonal direction, referred to as the "second wrapping event." Let $N_1$ denote the number of bonds at the first wrapping event, $N_2$ the number of bonds at the second wrapping event, and $N_e=L_1^2+L_2^2$ the total number of available edges on a system of size $[L_1, L_2]$. In the case of random percolation, where the wrapping probability is a smooth function of $n=N/N_e$ admitting the scaling collapse $\Pi_{wrap}(n)= f((n-n_c)L^{1/\nu})$, the wrapping interval $\Delta N = N_2 - N_1$ scales with system size as $\Delta N \propto L^{2-1/\nu}$~\cite{Newman2001}\footnote{{To be precise, Ref.~\cite{Newman2001} reports the wrapping probability as a function of the occupation probability $p$ by convolving ensembles at fixed rank $n$ with a binomial distribution. In this section we instead concern ourselves with ensembles at fixed $n$; see~\autoref{sect:random} below.}}, as does the standard deviation of $N_1$ and $N_2$ across samples. But in the checkerboard model with $\mathcal{L'}$ constructed from rational approximants to the Ford lattice (see~\appref{app:boundary} for the explicit construction), the behavior in the vicinity of $n_c$ is remarkably uniform across samples. For both sequences of system sizes, $\Delta N=1$. This naively yields $\nu=1/2$, but this spurious exponent is just a manifestation of the fact that $f$ in this case has a step function at the transition. To probe the critical region beyond this step, we avail ourselves of observables beyond the wrapping probability, discussed below.

Not only is $\Delta N=1$ in all samples, but the range of $N_1, N_2$ across samples is also of order 1. Explicitly, percolation instances can be divided into three groups. 
For system sizes of the form $[L_1^e, 0]$, $N_1=N_e/2$, $N_2 = N_e/2+1$; in the absence of two-fold anisotropy, the first wrapping event is equally likely to be in the $+$ or $-$ direction. Since the distribution of $b(x,y)$ is perfectly symmetric about 0, fixed $b=0$ ensembles all have a wrapping in only one direction.\footnote{This may seem to contradict our earlier statement that all samples in a given population have the same connected components. To be more precise, the bulk (non-percolating) clusters are identical. The wrapping cluster has the same composition of \textit{vertices} in each sample, but samples differ with respect to the placement of the "wrapping bond". In all samples, this percolating cluster spans the system (in the sense of open boundary conditions) in both directions, but only wraps around the periodic boundary in the $+$ or $-$ direction.} Samples with system sizes of the form $[L_1^o, L_1^o]$ fall into two categories. In the first, dubbed population I, the wrapping events are at strictly positive $b$,  with $(N_1, N_2) = (N_e/2+1, N_e/2+2)$. In the second, population II, the wrapping thresholds are strictly negative, $(N_1, N_2) = (N_e/2-1, N_e/2)$.

If we consider ensembles of fixed $b$ rather than fixed $n$, we find that on finite system sizes, a different spurious exponent is deduced, because $b(n)$ has several sharp steps near $n=1/2$, leading to qualitative differences in the scaling functions. This is discussed in further detail in~\appref{app:fss}. In the appendix, we also study the crossing probability with open boundary conditions and find that its scaling function is likewise dominated by steps very close to the critical point.

\subsection{Largest cluster ratio}\label{sect:dual}
Since the scaling collapse for both the wrapping and spanning probabilities is dominated by steps in $f(x)$ and in $b(n)$ very close to the critical point, we instead seek an observable whose scaling function is not just a step function. Our chosen observable is the ratio of the mass (number of sites) of the largest cluster $s_{max}$ at $n$ to that of the largest cluster at $1-n$:
\begin{equation}\label{eq:dual-log}
    R(n,n_c)= \langle \log [s_{max} (n)/s_{max}(2n_c-n)] \rangle
\end{equation}
where $n_c=1/2$, and the average is taken over fixed $n$ ensembles in the same population. This is statistically equivalent to taking the average of the log of the ratio of the mass of largest cluster on the original lattice to that on the dual lattice.

\begin{figure}[hbtp]
\subfloat[]{\includegraphics[width=0.5\linewidth]{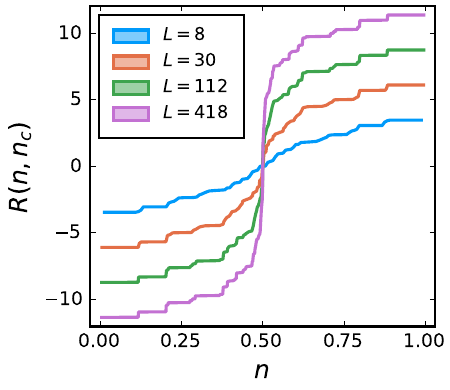}\label{fig:dual-unscaled}}
\subfloat[]{\includegraphics[width=0.5\linewidth]{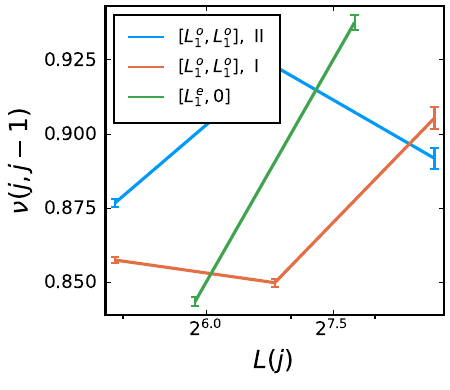}\label{fig:nu-checker-fit}} \\
\subfloat[]{\includegraphics[width=\linewidth]{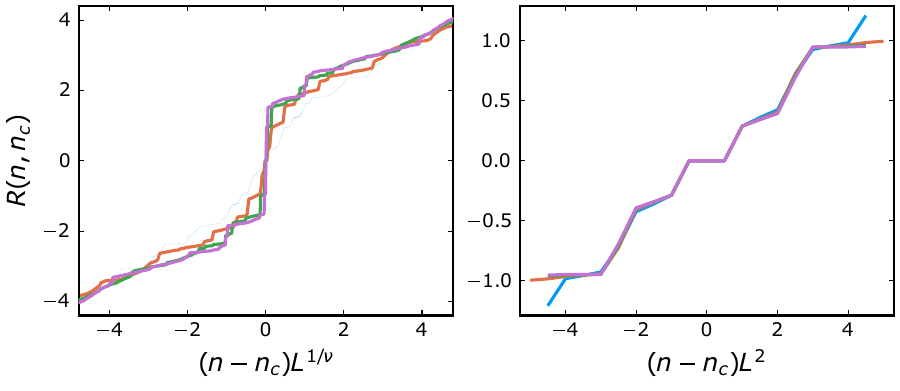}
    \label{fig:dual-scaled}}
\caption{(a) $R(n,n_c)$, as defined in~\autoref{eq:dual-log}, for population I of samples with $[L,L]$ PBCs in the checkerboard Ford model. (b) Critical exponent $\nu(j,j-1)$ inferred from the scaling of the slope of $R(n,n_c)$ for three populations (\autoref{eq:slope-nu-pbc}). (c) Scaling collapse of $R(n,n_c)$ for population I with $\nu=0.9$ (left, excluding $L=8$), and the zoomed-in spurious scaling collapse with $\nu=1/2$ (right).}
\end{figure}

Whereas the wrapping probability is strictly 0 or 1 outside the narrow interval around $n=n_c$, $R(n, n_c)$ is amenable to scaling over a larger interval because it increases monotonically all the way up to $n=1$ (\autoref{fig:dual-unscaled}). The scaling collapse to $R(n, n_c) = f((n-n_c)L^{1/\nu})$ is consistent with 
\begin{equation}\label{eq:nu-checker}
    \nu_{checker} \approx 0.9 \pm 0.1
\end{equation}
and is shown in the left panel of~\autoref{fig:dual-scaled} for population I of the odd parity sequence. 

 Overall, this method favors smaller $\nu$ than that inferred from the nominally infinite system methods, but has large uncertainties due to several factors. We can try to deduce $\nu$ by scaling the slope of $R(n,n_c)$ between consecutive system sizes $L(j)$, analogously to~\autoref{eq:slope-nu}:
\begin{equation}\label{eq:slope-nu-pbc}
    \nu(j,j-1) = \frac{\log[L(j)/L(j-1)]}{\log[m(j)/m(j-1)]}
\end{equation}
The slope $m(j)$ in the range $R(n,n_c)=[-R_0,R_0]$ is estimated as $R_0/(n_0-n_c)$ where $R(n_0,n_c)=R_0$. The inferred $\nu(j,j-1)$ turns out to be quite sensitive to $R_0$. With $R_0=3$, $\nu(j,j-1)$ converges fairly well with $j$ for the odd sequence of system sizes, $L=8,30,112,418$ (blue and orange curves in~\autoref{fig:nu-checker-fit}). However, a scaling collapse of comparable quality is obtained for slightly higher $\nu$, and the exponent deduced from even system sizes $L\sqrt{2}=22,82,306$ is not yet converging (green curve). 

Zooming in further, the scaling function for $R(n,n_c)$ also contains several small steps. Indeed, taking $R_0=1$, we would infer $\nu=1/2$, as from the wrapping probability. The scaling collapse with $\nu=1/2$ is shown in the right panel of \autoref{fig:dual-scaled}. Thus, the spurious exponent of $1/2$ arises from a rounding of the central step in the scaling function, which has $\nu\approx 0.9$, by sub-leading effects.

\begin{figure}[t]
    \centering
    \includegraphics[width=\linewidth]{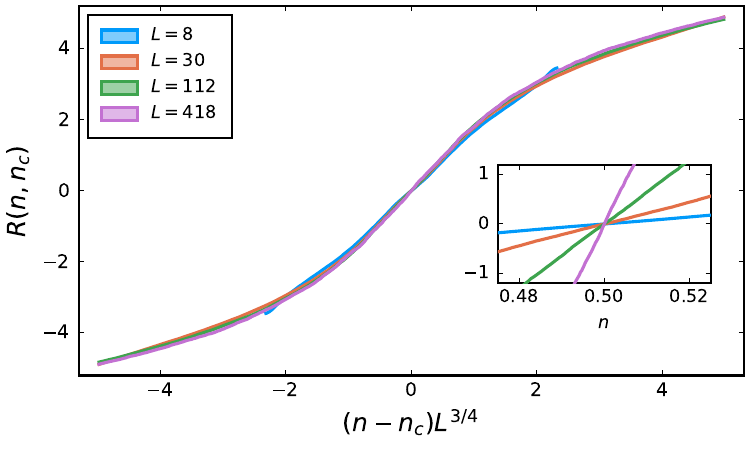}
    \caption{Scaling collapse of $R(n,n_c)$ with $[L,L]$ PBCs for random percolation, $\nu=4/3$, with the inset showing the smooth behavior of $R(n,n_c)$ (unscaled) zoomed in near $n_c=1/2$.}
    \label{fig:dual-random-zoom}
\end{figure}

In place of~\autoref{eq:dual-log}, we could instead define the log ratio by, rather than taking the ratio of the largest cluster size with $n_0$ and $N-n_0$ bonds on the original lattice, taking the ratio of the largest cluster size on the original lattice, to the largest cluster size on the dual lattice. Since populations I and II are related via a duality transformation, this alternate definition essentially relates a sample of population I at $n_0$, to a sample of population II at $N-n_0$. The main difference from our original definition is that, filtering on one of the populations, $R(n_c,n_c)$ is either strictly positive (for population I) or strictly negative (population II). The overall scaling remains consistent with our reported $\nu$.

From the scaling collapse with $\nu=0.9$, it would be tempting to conclude that $\nu<2/d=1$, which by the Harris criterion would imply that added randomness is relevant~\cite{Harris1974}. However, given the piecewise nature of the scaling function and the discrepancy between the exponent inferred from nominally infinite methods vs. finite size scaling, we do not have sufficient evidence to make this claim. Nevertheless, both methods of determining $\nu$ clearly exclude the random percolation exponent of $\nu=4/3$, for both the checkerboard and counter-rotated models.

\subsection{Comparison with random percolation}\label{sect:random}
In addition to having a different exponent $\nu$, the critical scaling of random percolation differs from the quasiperiodic Ford models in several ways. Here we offer a brief summary of these differences, some of which have already been mentioned in the foregoing discussion.

Random bond percolation is typically studied in the analog of fixed $b$ ensembles, where the deterministic parameter $b$ is replaced by the probability $p$ of occupying a bond. But the Newman-Ziff algorithm~\cite{Newman2000,Newman2001} naturally lends itself toward fixed $n$ ensembles, by assigning each bond a random "p-score" between 0 and 1, and adding bonds in order of increasing $p$. The typical wrapping threshold is then at $p_c=n_c=1/2$. As noted in Ref.~\cite{Ziff2010}, the two ensembles give similar results, unlike for the checkerboard model. A key difference from the checkerboard model is the absence of steps in either the scaling functions $f(x)$, $g(x)$ for any of the observables studied, or in the function $n(p)$, which just follows a binomial distribution at a given $p$. Therefore, scaling collapses of the form $f((n-n_c)L^{1/\nu})$ and $g((p-p_c)L^{1/\nu})$ are both consistent with the known exponent $\nu=4/3$. In particular, we obtain an excellent scaling collapse of $R(n,n_c)$ (\autoref{fig:dual-random-zoom}) with $\nu=4/3$, confirming that this is a valid, although nonstandard, observable for obtaining $\nu$. $R(n,n_c)$ is smooth on both large and small scales, as the average mass of the largest cluster increases continuously with $n$. A complementary method for obtaining $\nu$ comes from the scaling of the average wrapping interval, that is the difference in $n$ or $p$ between the first and second wrapping event. Whereas for the quasiperiodic checkerboard model all samples have the same wrapping interval $\Delta N=1$, for random percolation $\Delta N \propto L^{2-1/\nu}$, and $\Delta b \propto L^{-1/\nu}$, as expected.

The emergent conformal symmetry at the critical point of random percolation is well studied~\cite{Langlands1994,Smirnov2001,Duminil-Copin2020}. One consequence of this is that added two-fold anisotropy is marginal. The response to anisotropy in the checkerboard model is the focus of the next section.

\section{Two-fold anisotropy}\label{sect:anisotropy}
As a further demonstration of how quasiperiodicity manifests in a qualitatively different percolation transition from the random model, we also consider a perturbation away from square symmetry wherein the fractions of the bonds present along the two directions are different. %Having seen a lack of universality in~\autoref{sect:fractal}, for \gs{largely autobiographical reasons} in the remainder of the paper
{In this Section of our paper, we focus on the checkerboard model.}

Explicitly, let $q=(x+y) \mod 2$ denote the parity of the bond at position $(x,y)$; with our choice of origin, odd (even) bonds are oriented at an angle of $+\pi/4$ ($-\pi/4$) from horizontal.  Then the criterion for cutting this bond is modified from~\autoref{eq:model} to:
\begin{equation}\label{eq:anisotropy}
    b(x,y) > b + b'(-1)^q
\end{equation}
where $b'$ parameterizes the anisotropy between odd and even bonds. 

Once again, instead of working in terms of $b$, we can take fixed $(n_+, n_-)$ ensembles, where:
\begin{equation}
    n_\pm = n \pm n'.
\end{equation}
Choosing $n_+ \neq n_-$ breaks the symmetry of the original model under rotation by $\pi/2$.

For random percolation, such a 2-fold anisotropy is marginal: it results in a finite anisotropy in the scaling limit, so a line of fixed points with varying two-fold anisotropy. To wit, on the square lattice with probabilities $p_+$ and $p_-$ for occupying $\pm$ parity bonds, self-duality gives the critical line $p_+^c + p_-^c = 1$~\cite{Sykes1963,Sykes1964,Temperley1971}. But this entire fixed line is equivalent in the scaling limit to the isotropic system under a simple relative scaling of the two directions; if $p_+ = R p_-$ the crossover to quasi-1D percolation only occurs in the limits $R\rightarrow\infty$ and $R\rightarrow 0$~\cite{Redner1979}. This is \textit{not} the case in the checkerboard model; instead, the anisotropy is relevant and the phase transition breaks into two quasi-1D percolation transitions along the strong and weak axes, as sketched in~\autoref{fig:phase}. Here, we comment on some of the subtleties of this phase diagram, and the extent to which these features are unique to the Ford lattice checkerboard model.

\subsection{Fixed $n'$}
For quasiperiodic checkerboard percolation on the Ford lattice, we find that this added anisotropy in~\autoref{eq:anisotropy} is relevant. Explicitly, define the two-point function
\begin{equation}
    C(\vec{r}) = \langle \chi(\vec{v} \leftrightarrow \vec{v} + \vec{r}) \rangle
\end{equation}
where $\chi(\vec{v}\rightarrow \vec{u})$ denotes the event that vertices $\vec{v}$ and $\vec{u}$ are connected by a path of occupied bonds in a given sample. For $[L_1^e, 0]=[82,0]$ with PBCs, the average is taken over (1) all vertices $\vec{v}$ on the finite lattice (with $\vec{r}$ the displacement between $\vec{v}$ and $\vec{u}$ up to a periodic boundary lattice vector) and (2) 1200 samples. 

In the absence of applied bias, $C(\vec{r})$ has four-fold rotation symmetry (but not the full rotational invariance characteristic of random percolation at its critical point). But for $n'\neq 0$, the microscopic anisotropy induces perpendicular "strong" and "weak" axes along which the two-point function is respectively enhanced or reduced. Explicitly, we fix $n_\pm = 1/2 \pm n'$ and measure the two-point correlation function for all displacements $\vec{r}$. For $n'\neq 0$, intuitively one would expect the strong axis to align with the direction of the stronger bonds (i.e., oriented along angle $+\pi/4$ if $n'>0$), but we instead find that for $n'\gtrsim 0.015$, the two-point function is the most long-ranged along the horizontal direction, i.e. at an angle of $-\pi/4$ with respect to the applied bias, and shortest-ranged along the vertical direction. 

\begin{figure}[t]
\subfloat[]{
    \centering
    \includegraphics[height=0.45\linewidth,width=0.5\linewidth,keepaspectratio]{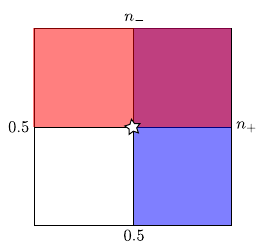}
    
    \label{fig:phase}}
\subfloat[]{
\includegraphics[width=0.5\linewidth,height=0.45\linewidth,keepaspectratio]{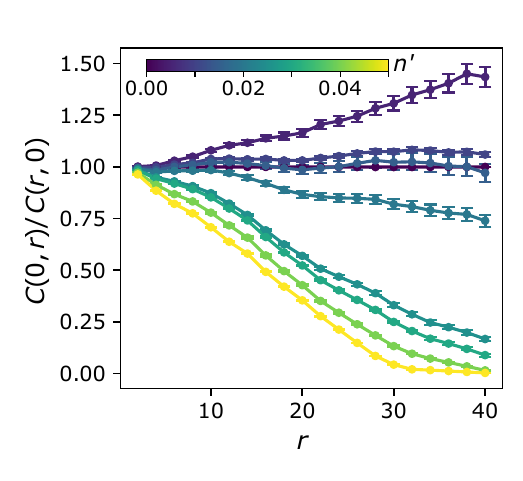}\label{fig:two-point}
}

\caption{ (a) Sketch of the two-parameter phase diagram for the checkerboard Ford model in the vicinity $0.3 < n_{\pm} < 0.7$ of the multicritical point. Blue (red) shading indicates the horizontally (vertically) percolating phase in the $n_+$ - $n_-$ plane.  [Farther from the multicritical point this phase diagram has more structure, but we have not thoroughly explored those regimes.] (b) Ratio of the two-point correlation function $C(\vec{r})$ along the vertical and horizontal axes, on a $[82,0]$ geometry with PBCs, for $n'=0$ (darkest blue), $0.005,0.01,0.015,0.02,0.25,0.03,0.04$, and $0.05$ (bright yellow), at $n=1/2$.\label{fig:anisotropy}}
\end{figure}

The ratio of the vertical to horizontal connectivity function $C(0,r)/C(r,0)$ is plotted as a function of the distance $r$ in~\autoref{fig:two-point}. For $n'=0$, all data points fall around $C(0,r)/C(r,0)=1$, owing to the four-fold symmetry. Notably, for $n'\gtrsim 0.015$ the ratio decreases as a function of $r$, rather than settling at a constant which could be scaled away. Finite-size effects prevent us from drawing strong conclusions about smaller $n'$; ~\autoref{fig:two-point} indicates that for [82,0] PBCs at $n'=0.005$ the strong axis is tilted toward the vertical, but this behavior is sensitive to the boundary conditions and the sample size. 

\subsection{Scaling collapse at $n_+=0.555$}
To further quantify the effect of the anisotropy, we fix $n_+$ and add $-$ bonds one by one in order of increasing $b(x,y)$, thus tracing a vertical line in the phase diagram of~\autoref{fig:phase}. Fixing $n_+$ in the range $0.3<n_+<0.7$ and scanning $n_-$, the percolation transition in the vertical direction remains at $n_-=1/2$ for $L\rightarrow\infty$.  For $0.5+O(1/L)<n_+<0.7$, the horizontal percolation transition, which is along the strong connectivity axis, occurs at lower occupation, $n_- \cong 0.293$. Thus we now have two percolation transitions, one for each direction. The scaling for both of these transitions is consistent with $\nu=1$, the exponent for 1D percolation.

\begin{figure}[t]
    \centering
        \subfloat[]{
    \includegraphics[width=\linewidth]{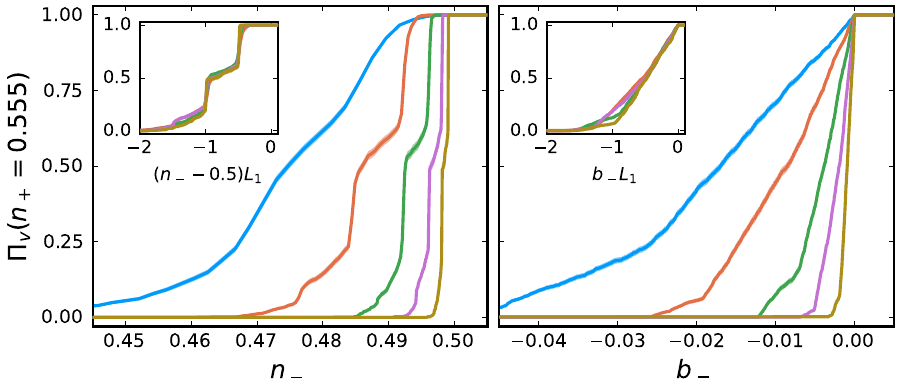}\label{fig:vert}} \\
    \subfloat[]{
    \includegraphics[width=\linewidth]{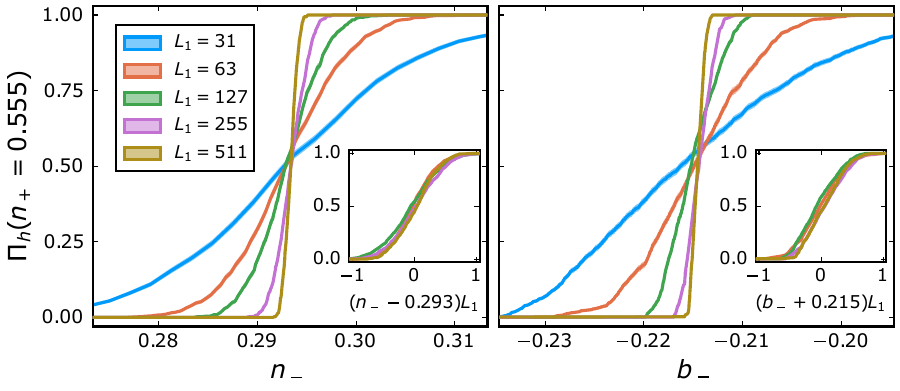}\label{fig:hor}}
    \caption{(a) Vertical crossing probability and (b) horizontal crossing probability in the checkerboard Ford model as a function of $n_-$ (left) and $b_-$ (right). All panels are at fixed $n_+=0.555$, with $\nu=1$ scaling collapse as an inset. $L_1=31$ (blue) is excluded from the scaling collapse.}
\end{figure}

In the isotropic case, scaling collapse of the wrapping probability with $b$ or with $n$ yields different exponents, both spurious, as elaborated upon in~\appref{app:fss}. Fortunately, this is not the case at a safe distance from the multicritical point; there the two scalings are both consistent with $\nu=1$ for this transition in the percolation along only one of the two directions. To wit, the scaling collapse for the vertical crossing probability (with open boundary conditions), $\Pi_v(n_+=0.555)$, is shown in~\autoref{fig:vert}, both in terms of $n_-$ (left) and $b_-$ (right). The scaling function in terms of $n_-$ consists of three steps, occurring at half-integer values of $L_1 (n_- - 0.5)$. As with the isotropic model, each step function has a slope $\propto 1/L^2$ for a finite system, so the curves with increasing $L$ are still sharpening up rather than collapsing neatly on top of each other. These steps are a consequence of the distribution of $b$ scores, which has discrete steps at finite system sizes; the scaling with $b_-$ is much smoother. But although the scaling functions $f$ and $g$ are quite different, it must be emphasized that both are consistent with $\nu=1$, and as the scaling collapse captures more than just the rounding of a single step function, this is a genuine critical exponent. Scaling of the quantity $0.5-\langle n_-^c\rangle(L) \propto L^{-1/\nu}$, where $\langle n_-^c\rangle(L)$ is the mean or median vertical crossing threshold at system size $L$, also yields $\nu=1$.

The vertical percolation transition occurs within the horizontally percolating phase, which is why different system sizes intersect at $\Pi_v=1$ in the unscaled plot. In turn, the horizontal percolation transition occurs at far lower $n_- \approx 0.293$, $b_- \approx -0.215$, as seen from the intersection of the two largest system sizes in~\autoref{fig:hor}. Since this transition occurs far away from the multicritical point where the steps in $b(n)$ are located, the scaling functions for $\Pi_h$ are smooth in terms of both $n_-$ and $b_-$.

We chose to use open boundary conditions here because with PBCs, only the even sequence of system sizes has boundaries aligned with the horizontal/vertical ($L_1^e=0$). It should be emphasized, though, that the data collapse with PBCs is also consistent with $\nu=1$, and the horizontal transitions occur at the same $b_-$ and $n_-$. Consistent results are also obtained from a scaling collapse of $R(n_-, 0.5)$ at $n_+=0.555$, even when we zoom in close to $n_-=0.5$. That is, while $R(n_-,0.5)$ still has steps near the transition like in the isotropic version of the model, these steps are consistent with $\nu=1$ rather than the spurious $\nu=1/2$ that arises in the isotropic case.

\subsection{Extended phase diagram}
\begin{figure}[t]
    \centering
    \subfloat[]{
    \includegraphics[width=0.45\linewidth,height=0.5\linewidth,keepaspectratio]{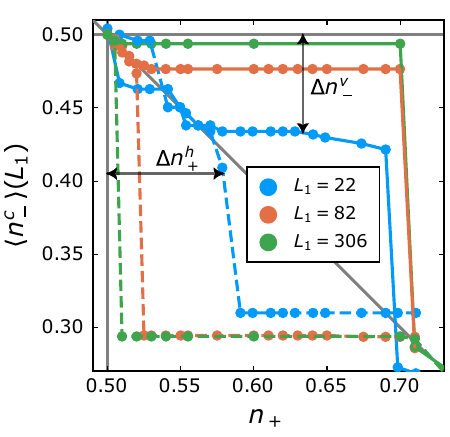}
    \label{fig:varying-nplus}
    }
    \subfloat[]{
    \includegraphics[width=0.5\linewidth,height=0.45\linewidth,keepaspectratio]{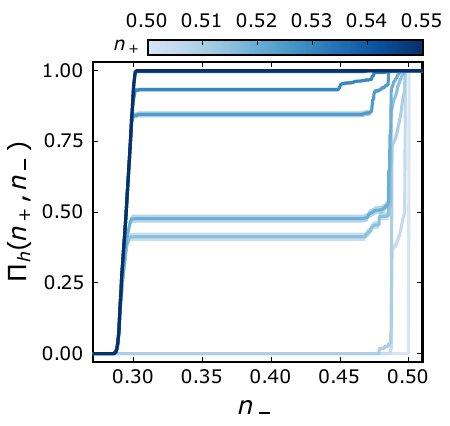}
    \label{fig:horizontal-steps}
    }
    \caption{(a) Median wrapping thresholds in the vertical (solid lines) and horizontal (dashed lines) direction as a function of $n_+$ in the checkerboard Ford model. Gray vertical and horizontal lines indicate the putative thresholds for infinite $L$, and $\Delta n_+^h$, $\Delta n_-^v$ are the deviations of the finite $L$ thresholds from these respective phase boundaries. Gray line of slope $-1$ is the line of varying $n'$ that passes through the multicritical point. (b) Horizontal wrapping probability for $n_+$ ranging from $0.5$ to $0.55$, for $[82,0]$.}
    \label{fig:phase-v2}
\end{figure}

The full phase diagram in the vicinity of the multicritical point can be completed using the underlying symmetries of the model. We now elaborate further on the following hypothesized scenario: for $L\rightarrow \infty$, the multicritical point immediately split into horizontal and vertical phases, and the horizontal axis is the strong axis for arbitrarily small $n'$. 
To see why this is sensible based on the data at finite sizes, let $\Delta n_-^v(n_+,L_1) = 1/2-\langle n_-^c \rangle_v (L_1)$ denote the deviation of the median vertical wrapping or crossing threshold below infinite system size limit of $(n_-^c)_v=1/2$, at a fixed $n_+$. This quantifies the difference between the phase boundary in a finite sample and the putative phase boundary drawn in~\autoref{fig:phase}. Similarly, let $\Delta n_+^h(n_-, L_1)$ denote the deviation of the finite system size horizontal percolation threshold and the hypothesized phase boundary at $n_-=0.5$ extending down to $n_- = (n_-^c)_h\approx 0.293$. These quantities are indicated in~\autoref{fig:varying-nplus}. As noted in the previous subsection, $\Delta n_-^v(0.555,L)\propto 1/L$, a relation that approximately holds throughout the interval $0.5 < n_+ < 1-(n_-^c)_h$. Under the transformation from the original lattice to the dual lattice, $n_+\leftrightarrow 1-n_-$, $\Pi_v \leftrightarrow 1-\Pi_h$, so we expect---and~\autoref{fig:varying-nplus} confirms---that $\Delta n_+^h$ also scales as $1/L_1$. Thus, as $L$ increases, the interval in $n_+$ in which the horizontal wrapping threshold drops from $1/2$ to $(n_-^c)_h$ shrinks toward zero.

Now let's return to the setup in which we fix $n_\pm = 1/2 \pm n'$ and vary $n'$, which corresponds to a line with slope $-1$ through the multicritical point. The ratio of the two-point connectivity function along the vertical and horizontal, plotted for $[82,0]$ in~\autoref{fig:two-point}, presents a puzzle: for very small positive $n'$, the two-point function is in fact \textit{greater} along the vertical axis, seemingly contrary to our proposed phase diagram. As seen from~\autoref{fig:varying-nplus}, however, for finite $L$ the line $n_+ + n_- = 1$ passes through a critical region at small $n'$ where the horizontal and vertical percolation transitions cannot be cleanly separated, and where for some samples the vertical wrapping cluster develops first. Thus whereas for sufficiently large $n'$ all samples are in the horizontally percolating, vertically non-percolating phase, below an $L$-dependent cutoff $n'_c(L)$, the opposite can occur. For individual samples, this critical region can extend to somewhat larger $n'$ than~\autoref{fig:varying-nplus} suggests, as the distribution of horizontal wrapping thresholds becomes bimodal with a peak near $(n_-^c)_h$ and another near $1/2$, before changing to a single peak at $(n_-^c)^h$ for sufficiently large $n_+$ (\autoref{fig:horizontal-steps}). While we cannot definitively rule out an alternate phase diagram where $n'_c(L\rightarrow\infty)$ remains finite, the trend with increasing $L$ suggests that this critical region shrinks toward the multicritical point as $L\rightarrow\infty$, yielding the phase diagram advertised in~\autoref{fig:phase}. A similar phase diagram can be drawn in terms of $b_+$ and $b_-$, but for a finite sample, the finite size effects at small $b'$ will become entangled with the presence of sharp steps in the function $b'(n')$.

Since the horizontal percolation phase boundary extends down to $n_-= (n_-^c)_h\approx 0.293$ for a wide range of $n_+>1/2$, the self-duality of the lattice implies that the vertical phase boundary remains at $(n_-^c)_v=1/2$ out to $n_+\approx 0.707$. Beyond this point, as shown in~\autoref{fig:varying-nplus}, the vertical wrapping threshold decreases sharply. In this regime, the separation into horizontally and vertically percolating phases is no longer appropriate, as the principal axes of the anisotropy begin to rotate away from the horizontal/vertical and toward the axes of the applied bias. The results for $n'>0.7$ depend on the system size and boundary conditions, and we have not carefully explored that part of this phase diagram. 

\subsection{Comparison with other models}
One of the many surprises of the Ford checkerboard model is that in the presence of two-fold anisotropy, the strong/weak axes align with horizontal/vertical, whereas the average two-point connectivity along the axes of applied bias remain equal. This is in contrast to random percolation, where for large enough $n'$ $C(\vec{r})$ is larger by a constant factor along the microscopically favored direction ($\pi/4$). The orientation of the strong axis also depends on the lattice $\mathcal{L'}$ used in the checkerboard construction. While two-fold anisotropy appears to be relevant for $\vec{a} = (37805/46962,1/2)$ found via the optimization protocol, %the phase diagram in that case is more complicated. We find a larger interval in $n'$ over which the strong axis is rotating (perhaps another finite-size effect like that discussed above), and 
at large bias, the strong axis aligns at an angle somewhat below $\pi/4$. 

A special feature of the Ford checkerboard model that explains, in part, the preference for the horizontal/vertical axes is 
the strong sensitivity of the "local occupation rates" to arbitrarily weak bias. Consider a lattice with periodic boundaries aligned with $[L_1,0]$ and $[0,L_1]$, so that the number of available bonds in each row and column is $L_1$, half of which are $+$. For $x_0=0,...,L_1-1$, the quantity
\begin{equation}
    n_\pm(x_0) = \frac{|\mathcal{E} \cap \{(x,y)\in \mathcal{L}: x=x_0\}|}{L_1/2}
\end{equation}
is the occupation rate of $\pm$ bonds in the column with $x$ coordinate $x_0$. 

\begin{figure}[t]
    \centering
    \subfloat[]{
    \includegraphics[width=\linewidth]{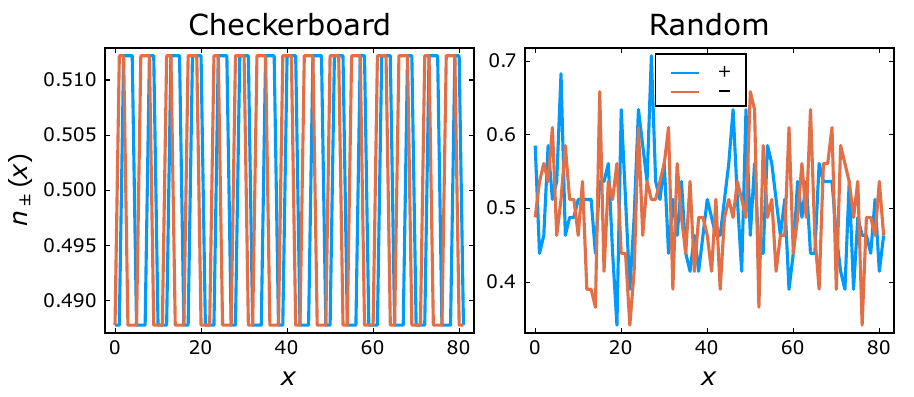}\label{fig:pbc-local}
    } \\
    \subfloat[]{
    \includegraphics[width=\linewidth]{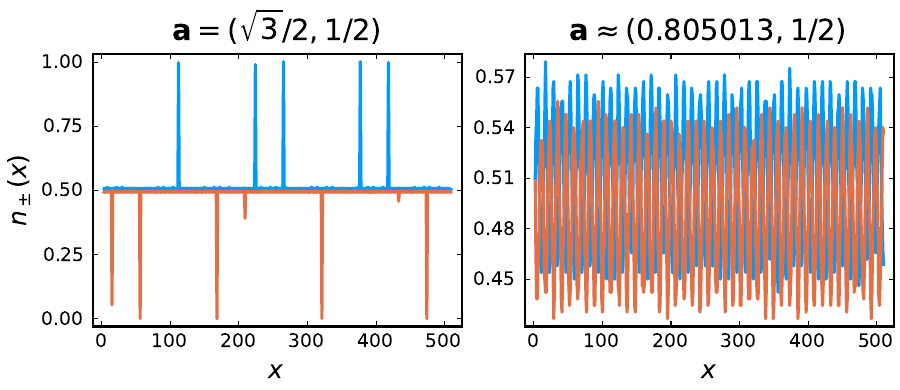}
    \label{fig:obc-local}
    }
    \caption{Local occupation rates of $+$ and $-$ bonds along columns of the lattice. (a) Occupation rates for system size [82,0] with PBCs at $n_+=n_-=1/2$ for quasiperiodic checkerboard Ford (left) and random percolation (right). (b) Occupation rates for system size [511,0] with OBCs at $n_+=0.51, n_-=0.49$ for checkerboard Ford (left) and the quasiperiodic checkerboard model with optimized parameters (right).}
    \label{fig:local-rates}
\end{figure}

As shown in~\autoref{fig:pbc-local}, at a global occupation rate of $n_+=n_-=1/2$, the local occupation rates in an individual sample of the Ford model are very homogeneous: $N_\pm(x)$ oscillates quasiperiodically between $L_1/2+1$ and $L_1/2-1$. In comparison, the local occupation rates for a random percolation instance are statistically homogeneous, but with a significantly larger variance. The same trend holds with open boundary conditions, excluding the columns along the edges. Thus, at precisely $n_+=n_-=1/2$ the quasiperiodic Ford model is homogeneous in a much stronger sense than the random model.

This changes dramatically when we turn on the anisotropy: for $n'>0$, there are rare columns where $n_-(x)=0$, and others where $n_+(x)=1$. This is shown for $n'=0.01$ and open boundary conditions in~\autoref{fig:obc-local}. 
In~\autoref{fig:obc-local} we also plot the local occupation rates at the same $n'$ for $\vec{a}=37805/46962,1/2)$. In this case, the distributions of $n_\pm(x)$ have significant overlap, and both distributions become narrower as $L$ increases, while also remaining narrower than the distributions in the random percolation model at the same $n'$. For some purposes, where we might wish to avoid the static 1d inhomogeneity in local occupation rates found in the checkerboard Ford model, this makes the optimized choice of $\vec{a}$ appealing.

Another alternative to the Ford checkerboard model which sidesteps some of these peculiarities is the counter-rotated model analyzed in previous sections. The checkerboard model is able to "select" strong/weak axes other than the axes $\pm \pi/4$ of microscopic bias because it lacks any reflection symmetry. To wit, note that taking $\mathcal{L'}$ to be the Ford lattice with $\vec{a}_-=(\sqrt{3}/2, -1/2)$ in place of $\vec{a}_+=(\sqrt{3}/2,1/2)$ would swap the behavior along the horizontal and vertical axes in the above discussion, making the vertical axis the strong connectivity axis for $n'>0$. On the other hand, one of the advantages of the counter-rotated model is its possession of additional reflection symmetries. The two-fold anisotropy breaks the symmetry under reflection through the horizontal and vertical axes but preserves the reflection axes $\pm \pi/4$. This implies that the strong and weak axes must align with the microscopic bias, along $\pm \pi/4$, like in random percolation. The relevance of two-fold anisotropy in this model remains a question for future work.
\section{Discussion}\label{sect:discuss}
The quasiperiodic models discussed in this paper have several features that distinguish them from random bond percolation, including: critical exponent $\nu < 4/3$, a self-similar sequence of four-fold or two-fold symmetric fractal clusters with different fractal dimension than random critical percolation, and relevance of two-fold anisotropy.  {Thus replacing randomly chosen bonds with deterministically and quasiperiodically chosen bonds changes the universality class of the percolation transition.  This is in contrast with the three-dimensional Anderson localization transition for noninteracting particles, where such a change from random to quasiperiodic potential does not change the universality class~\cite{Devakul2017}}.

Next we briefly mention some open questions that remain for future study:

Further research is required to determine whether there exists a larger universality class of quasiperiodic models, or whether the critical behavior is specific to the choice of $\mathcal{L'}$ and other specifics of the models. The latter alternative is true of variations on the counter-rotated model, as evidenced by the changing properties of fractal clusters and hulls as we tuned the parameter $a$. However, it may still be the case that a more robust universality class does exist for quasiperiodic models with a different symmetry.

In this paper, we focused on just one member of the checkerboard and counter-rotated classes of models, the Ford lattice. Modifying $\vec{a}$ changes the fractal sequence of cluster sizes and the specifics of the response to two-fold anisotropy, although the estimated $\nu$ for a checkerboard model using one of the optimized choices of $\mathcal{L'}$ ($\vec{a}\approx(0.805013,1/2)$) is roughly consistent with that obtained here. Thus, the invention and exploration of other substantially different quasiperiodic percolation models (such as those of Ref.~\cite{Chernikov1994}) will be interesting, to see what variety of behavior can occur.
%It would also be informative to complete the Ford lattice phase diagram by analyzing the anisotropic phases further away from the multicritical point.
 %\subsection{Conformal invariance}
%The relevance of two-fold anisotropy is a strong indication that the multicritical point is not conformally invariant, unlike the critical point for random percolation.  
Could there be some other nonrandom quasiperiodic percolation models that, like random percolation, do have an emergent conformal invariance at criticality, or is that not possible without randomness? 
In a conformally invariant finite system, the crossing probability between the intervals $[x_1, x_2]$ and $[x_3, x_4]$ on opposite boundaries, which can be expressed in terms of four-point functions of boundary operators, depends only on the cross-ratio, $(x_4-x_3)(x_2-x_1)/(x_3-x_1)(x_4-x_2)$~\cite{Cardy1992}. In future studies, it would be useful to verify explicitly the failure of this ansatz for checkerboard and counter-rotated models.
 
As mentioned in the introduction, our motivation for this work comes from the application to monitored quantum circuits.
As a preliminary study in this direction, we considered a spacetime translation-invariant Clifford circuit with dual-unitary gates which is a "good scrambler" in terms of entanglement generation and contiguous code length in the absence of measurements~\cite{Sommers2022b}. When {projective measurements are added to this circuit in a quasiperiodic pattern} according to the checkerboard prescription, we find that there is a measurement-induced phase transition far from the self-dual point, at $b\approx -0.306$ ($n\approx 0.225$). While follow-up work is necessary to determine the critical exponents, the present evidence suggests $\nu\approx 1$, which falls outside the universality class of the random Clifford transition. This is in contrast with the findings of Ref.~\cite{Li2019}, which used a different quasiperiodic arrangement of measurements defined by a superlattice whose lattice vectors are aligned with the underlying circuit, and found {approximately} the same exponents as the random model. The {random Clifford universality class of the measurement-induced transition} has emergent conformal invariance~\cite{Li2020}, but is distinct from the random percolation universality class~\cite{Zabalo2022}. Thus, in future work it will be interesting to compare the exponents of the quasiperiodic percolation transition {to those of the} quasiperiodic circuit transition.  Is there any sense in which we can define "spacetime clusters" at the circuit critical point with discrete scale invariance, as at the percolation threshold?

\begin{acknowledgements}
We thank Peter Sarnak for introducing us to the Ford lattice and Romain Vasseur and Sarang Gopalakrishnan for helpful discussions.  Work supported in part by NSF QLCI grant OMA-2120757. GMS is supported by the Department of Defense (DoD) through the National Defense Science \& Engineering Graduate (NDSEG) Fellowship Program.
Numerical work was completed using computational resources managed and supported by Princeton Research Computing, a consortium of groups including the Princeton Institute for Computational Science and Engineering (PICSciE) and the Office of Information Technology's High Performance Computing Center and Visualization Laboratory at Princeton University.
\end{acknowledgements}
\appendix
\counterwithin{figure}{section}
\counterwithin{table}{section}
\section{More details on fractal clusters}\label{app:fractals}
In the checkerboard model, there are two species of fractal clusters, whose discrete scale invariance and four-fold (not continuous) rotational invariance means that the critical point is not conformally invariant. Meanwhile, the critical clusters in the counter-rotated model have only two-fold symmetry, with clusters of a given size coming in pairs related by reflection. In this section, we provide more details on the structure of these clusters and the corresponding hulls. Although we have not been able to arrive at an analytic form for the sequence of cluster sizes, their rich structure is closely tied to the discrete scale-invariant pattern of near-commensurate points on the Ford lattice.

\subsection{Checkerboard model}
\begin{figure*}
\centering
\includegraphics[width=0.75\linewidth]{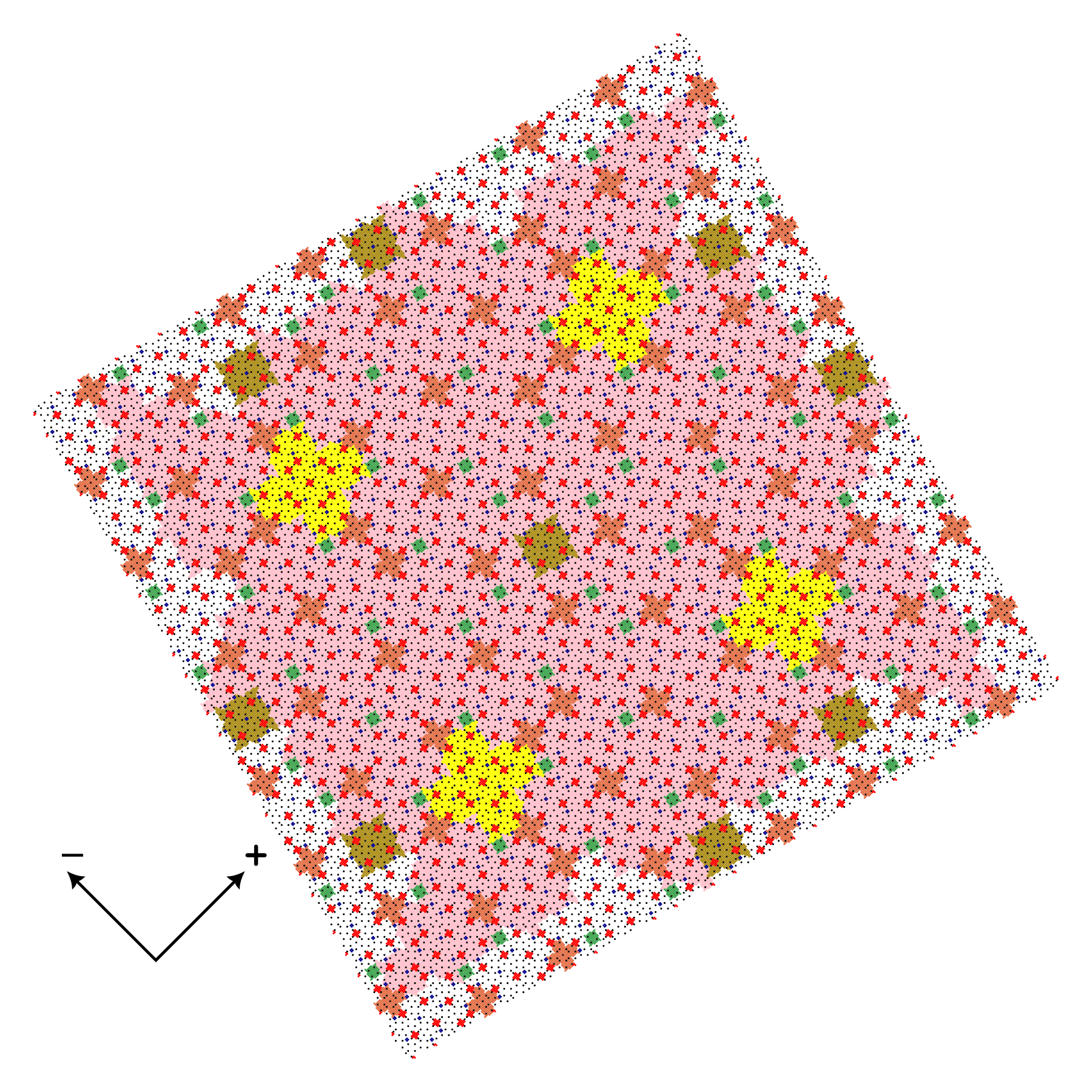}
\includegraphics[width=0.5\linewidth]{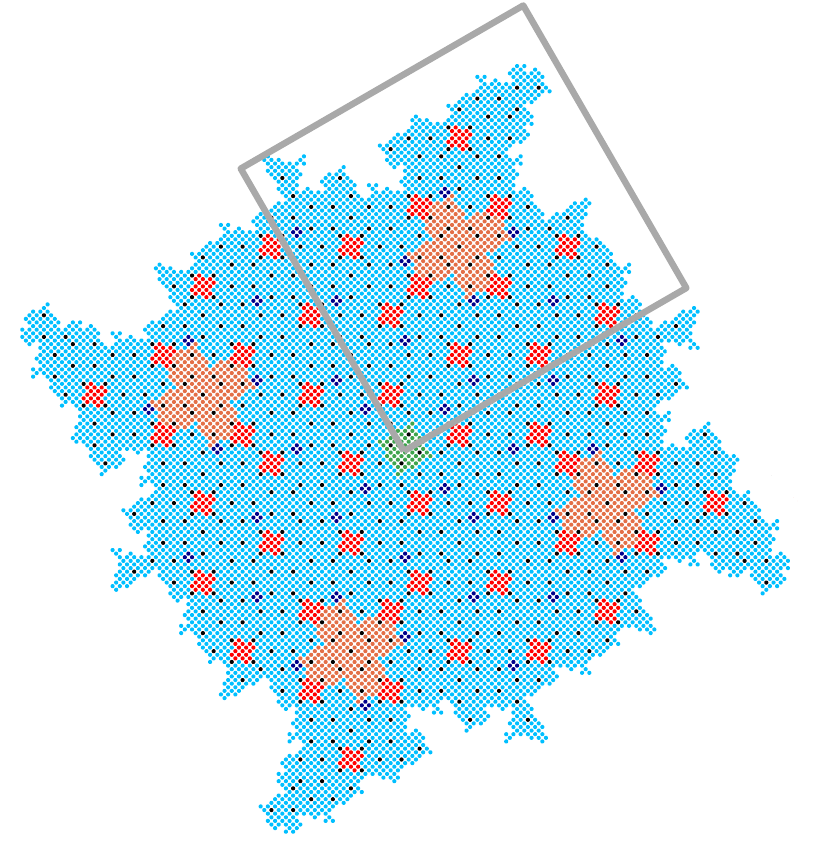}
\includegraphics[width=0.3\linewidth]{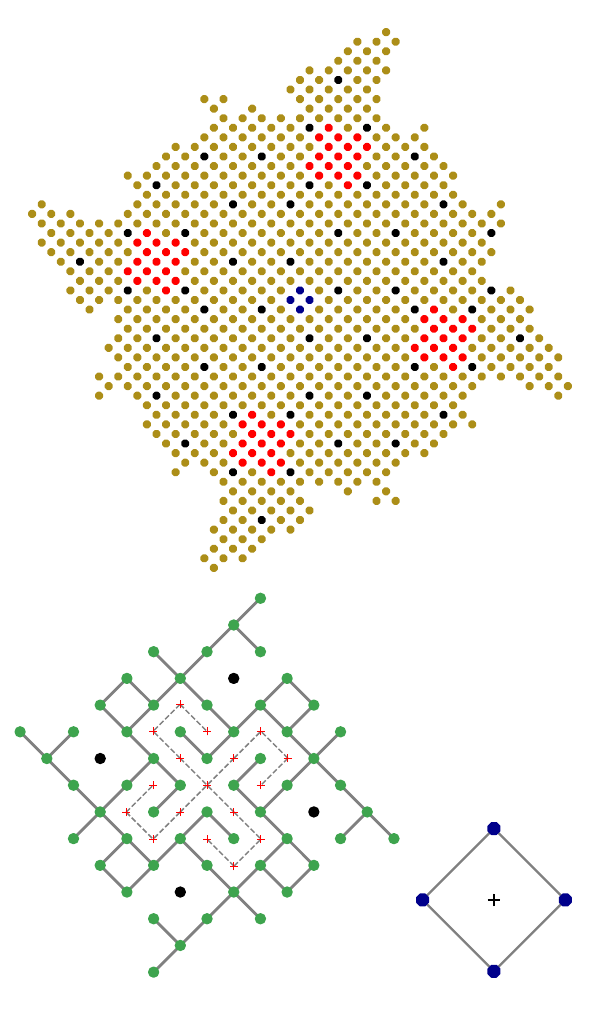}
\caption{Species A of fractal clusters in the checkerboard Ford model: mass 107064 (pink, with smaller clusters around its periphery also shown), 8632 (blue, with the gray box indicating the quadrant shown in~\autoref{fig:fractal}), 696 (gold), 56 (green), and 4 (dark blue). For the two smallest clusters, the edges are also shown, and the clusters (of mass 17 and 1 respectively, belonging to species 2) enclosed on the dual lattice are indicated with dashed lines and +'s.  {The arrows labelled $-$, $+$ indicate the bond directions in the underlying lattice $\mathcal{L}$. }\label{fig:pictures-a}}
\end{figure*}

\begin{figure*}
\centering
\includegraphics[width=0.75\linewidth]{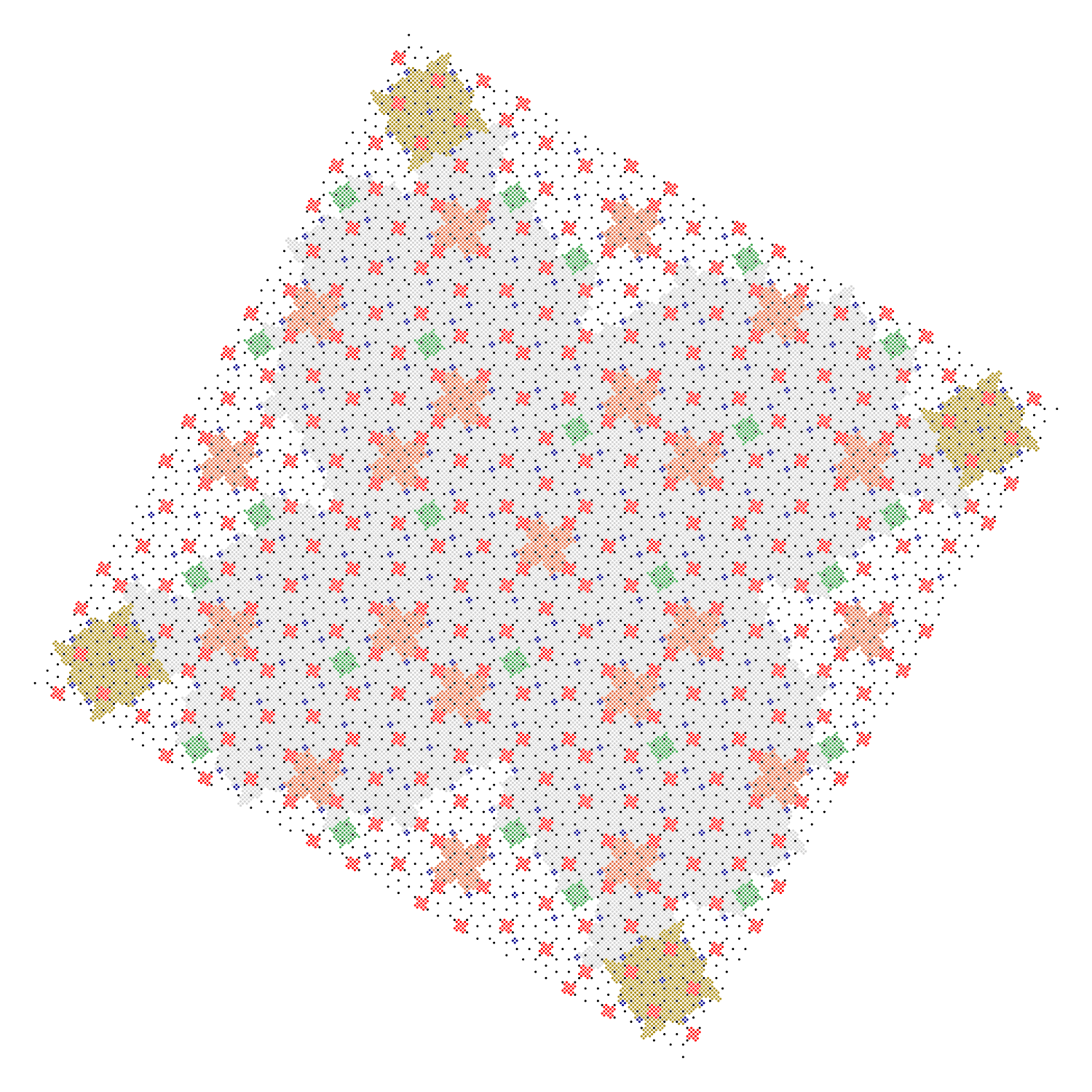}
\includegraphics[width=0.5\linewidth]{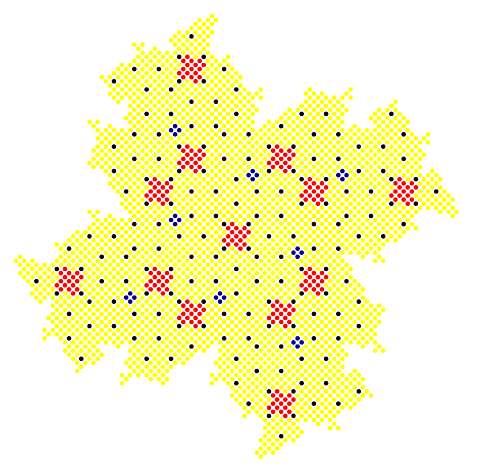}
\includegraphics[width=0.3\linewidth]{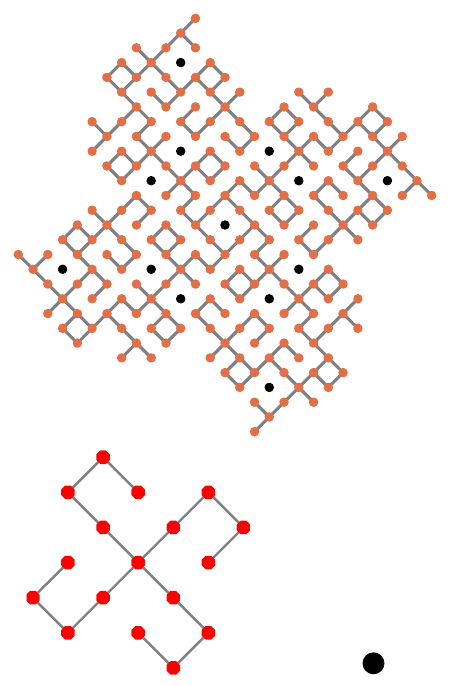}
\caption{Species B of fractal clusters in the checkerboard Ford model: mass 31952 (gray, with smaller clusters around its periphery also displayed), 2576 (yellow), 208 (orange), 17 (red), and 1 (black). For the three smallest clusters, the edges contained in the cluster are also shown. \label{fig:pictures-b}}
\end{figure*}

\autoref{fig:pictures-a} and~\autoref{fig:pictures-b} show the first five generations of the A and B species in the checkerboard model. These figures were obtained by applying the incremental percolation method with periodic boundary conditions, for which all the clusters on the finite system are generated (rather than just growing a single cluster). At $b=b_c$, $n=n_c$, all of the clusters that do not wrap around the boundary are precisely the ones identified in the nominally infinite system, and constitute the discrete steps in the cluster size distribution seen in~\autoref{fig:cumul}.

Clusters belonging to species A are rotated squares with small ornamental features. As $j\rightarrow\infty$, the edges of the square enclosing the cluster become aligned with the principal lattice vectors of the Ford lattice, $\vec{a}=(\sqrt{3}/2,1/2)$ and $\vec{a}^\perp = (-1/2,\sqrt{3}/2)$. Species B consists of "pinwheel"-shaped clusters, and the edges of the enclosing square align with $(\sqrt{3}/2,-1/2)$ and $(1/2,\sqrt{3}/2)$ as $j\rightarrow\infty$, which are the lattice vectors of a counter-rotated Ford lattice at $\theta=-\pi/6$. Thus, whereas the underlying lattice $\mathcal{L}$ has the point group $D_4$, with axes of reflection symmetry along the horizontal, vertical, and $\pm \pi/4$~\cite{Ashcroft1976}, superimposing the checkerboard at the Ford angle of $\theta=\pi/6$ preserves only the four-fold rotational symmetry. The fractal clusters exhibit no additional emergent symmetry at the critical point.

Since members of both species are four-fold symmetric, the inertia tensor is diagonal in any basis. The mass of each cluster scales approximately as $M(j)\propto r(j)^{D_f}$ where $r$ is the radius of gyration. In the main text, $D_f$ was determined from the scaling of the mass vs. radius of gyration of the first seven generations on a nominally infinite lattice. On finite system sizes of length $L$, $D_f$ can also be inferred from the mass of the largest cluster, which scales as $s_{max}(L) \propto L^{D_f}$ at the critical point, and from the average cluster size, which is the second moment of the cluster size distribution and scales at the percolation threshold as:
 \begin{equation}
     \langle s^2 \rangle(L) \propto L^{2-\eta}
 \end{equation}
where the exponent $\eta$ is related to the fractal dimension via the hyperscaling relation $\eta=2+D-2D_f$~\cite{Stauffer1979}. Consistent with the hyperscaling relation, the estimates of $D_f$ inferred from the scaling of $s_{max}$ and $\langle s^2\rangle$ converge to the same value. This value is in agreement with~\autoref{eq:df-checker}, but with less precision since the computational burden of finding all the clusters, rather than just growing a single cluster, limits the accessible system sizes. 

Ideally, we could determine $D_f$ exactly by finding an analytic expression for the sequences $A(j)$ and/or $B(j)$. We leave this as a challenge to the game reader and compile the details of the first seven generations in~\autoref{table:a} and~\autoref{table:b} for species A and B respectively.

\begin{table*}[t]
\centering
\begin{tabularx}{\linewidth}{|l|X|X|X|X|X|X|X|}
\hline
--- & $V$ & $E$ & $E-V$ & Dual & $N_{hull}$ & $N_{in}$ & $N_{out}$ \\ \hline
1 & 4 & 4 & 0 & 1 & 12 & 4 & 8\\\hline
2 & 56 & 60 & 4 & 17& 140 & 48 & 64\\\hline
3 & 696 & 756 & 60 & 208 & 1500 & 508 & 560 \\\hline
4 & 8632 & 9388 & 756 & 2576 & 16092 & 5368 & 5560\\\hline
5 & 107,064 & 116,452 & 9388 & 31952 & 172,652 & 57284 & 57936 \\\hline
6 & 1,327,928 & 1,444,380* & 116,452* & 396,304* & 1,852,396 & 613,304 & 615,640\\\hline
7 & 16,470,456 & 17,914,836* & 1,444,380* & 4,915,408* & 19,874,492 & 6,575,668 & 6,583,864\\\hline
$j$ & $A(j)$ & $E_A(j)$ & $E_A(j-1)$ & $B(j)$ & $N_{hull}^A(j)$ & $N_{in}^A(j)$ & $N_{out}^A(j)$\\\hline
\end{tabularx}
\caption{First nine generations of critical clusters of species A in the checkerboard Ford model. The columns from left to right are: the generation $m$; number of vertices $V$; number of edges $E$; $E-V$; the mass of the largest cluster enclosed by this cluster on the dual lattice; number of links in the external hull $N_{hull}$; number of occupied edges adjacent to the hull $N_{in}$; and number of unoccupied edges adjacent to the hull $N_{out}$. Entries marked with an asterisk are conjectured based on the trend observed in the first five generations. \label{table:a}}
\end{table*}
\begin{table*}[t]
\centering
\begin{tabularx}{\linewidth}{|l|X|X|X|X|X|X|X|}
\hline
--- & $V$ & $E$ & $E-V$ & Dual & $N_{hull}$ & $N_{in}$ & $N_{out}$  \\ \hline
1 & 1 & 0 & -1 & --- & --- & --- & --- \\\hline
2 & 17 & 16 & -1 & --- & 68 & 16 & 28\\\hline
3 & 208 & 224 & 16 & 4 & 756 & 224 & 264 \\\hline
4 & 2576 & 2800 & 224 & 56 & 8116 & 2584 & 2716 \\\hline
5 & 31952 & 34752 & 2800 & 696 & 87076 & 28448 & 28904\\\hline
6 & 396,304 & 431,056 & 34752* & 8632* & 934,244 & 307,736 & 309,348\\\hline
7 & 4,915,408 & 5,346,464* & 431,056* & 107,064* & 10,023,572 & 3,310,808 & 3,316,496\\\hline
$m$ & $B(j)$ & $E_B(j)$ & $E_B(j-1)$ & $A(j-2)$ & $N_{hull}^B(j)$ & $N_{in}^B(j)$ & $N_{out}^B(j)$\\\hline
\end{tabularx}
\caption{First seven generations of critical clusters of species B in the checkerboard Ford model, with columns as in~\autoref{table:b}. In the column labeled "Dual," --- indicates that the cluster encloses no clusters on the dual lattice (i.e., is a tree). \label{table:b}}
\end{table*}

\begin{figure}[hbtp]
    \centering
    \includegraphics[width=\linewidth]{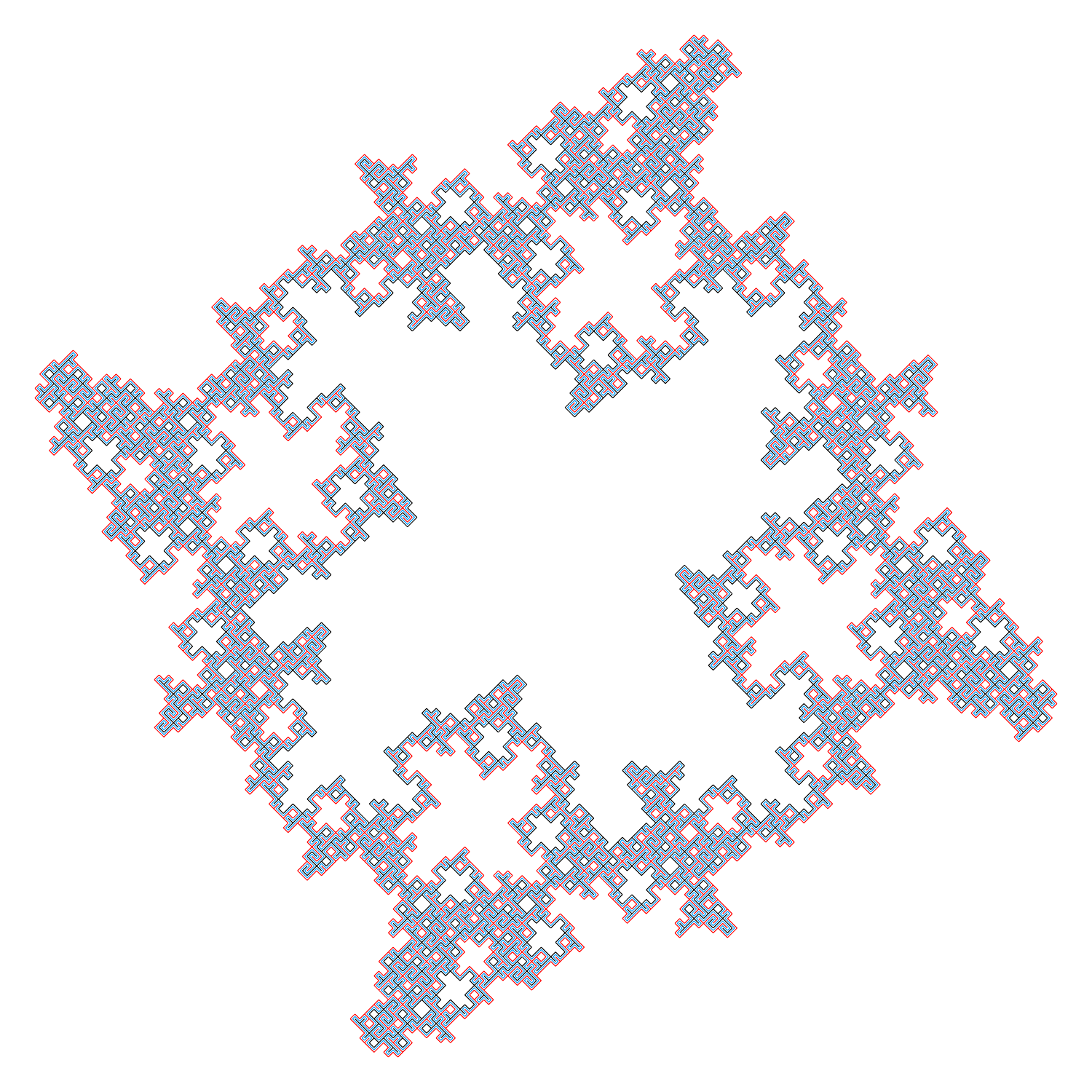}
    \includegraphics[width=\linewidth]{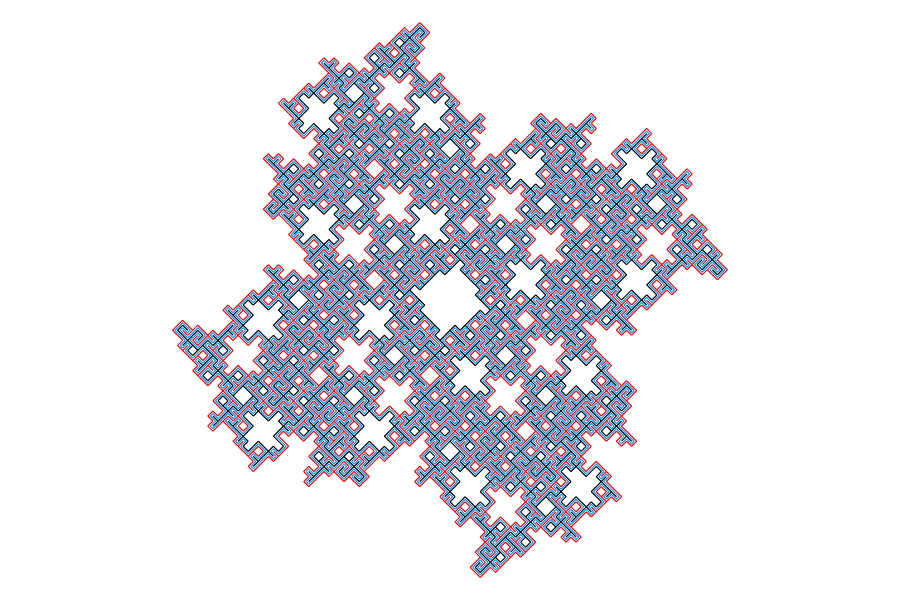}
    \caption{External perimeter of the fourth-generation A cluster ($M=8632$, $N_{hull}=16092$, top) and fourth-generation B cluster ($M=2576, N_{hull}=8116$, bottom) in the Ford checkerboard model. As in~\autoref{fig:counter-hull}, black, red, and blue lines respectively indicate the inner bonds, outer bonds, and "hull itself".}
    \label{fig:checker-hull}
\end{figure}

As seen in~\autoref{fig:pictures-a} and~\autoref{fig:pictures-b}, each cluster contains holes occupied by younger generations of each species. Scaling up from generation $m$ to $m+1$, the holes inside the cluster also scale up by one generation, and new holes appear belonging to the youngest generation. Moreover, examining the edges belonging to each cluster also allows us to determine the composition of clusters on the dual lattice. The $m$th generation of species A encloses the $m$th generation of species B on the dual lattice. In turn, for $m>2$, the $m$th generation of species B only encloses a much smaller cluster on the dual lattice, namely the $(j-2)$th generation of species A. This is related to the fact that the external perimeters of B clusters have a much smaller "inner core" than those of A clusters (\autoref{fig:checker-hull}). Large enough generations also enclose several smaller clusters on the dual lattice. Explicitly, a graph containing $E$ edges and $V$ vertices encloses $E-V+1$ dual lattice clusters. Here another facet of the fractal structure emerges: the quantity $E-V$ for the $m$th generation is equal to the number of edges for the $(j-1)$th generation. In summary, then, the two species are closely related both on the original lattice and on the dual lattice.
\begin{table}[t]
\centering
\begin{tabularx}{\linewidth}{|l|X|X|X|X|X|}
\hline
--- & $L$ & $V$ & $N_{hull}$ & $N_{in}$ & $N_{out}$ \\ \hline
1 & 1 & 2 & 8 & 1 & 6 \\ \hline
2 & 6 & 15 & 44 & 16 & 24 \\ \hline
3 & 28 & 132 & 244 & 108 & 118 \\ \hline
4 & 110 & 1244 & 1420 & 654 & 664 \\ \hline
5 & 416 & 11780 & 8276 & 3836 & 3846 \\ \hline
6 & 1558 & 111,580 & 48236 & 22382 & 22392 \\ \hline
7 & 5820 & 1,056,900 & 281,140 & 130,476 & 130,486 \\ \hline
8 & 21726 & 10,011,100 & 1,638,604 & 760,494 & 760,504 \\ \hline
9 & - & - & 9,550,484 & 4,432,508 & 4,432,518 \\ \hline
$m$ & $L(j)$ & $V(j)$ & $N_{hull}(j)$ & $N_{in}(j)$ & $N_{in}(j)+10$ \\\hline
\end{tabularx}
\caption{First nine generations of critical clusters in the counter-rotated model. $L$ is the side length of a square enclosing the cluster (which is oriented at $\pm \pi/4$ with respect to the square), and $V$ is the number of sites. $N_{hull}$, $N_{in}$, $N_{out}$ are the number of hull links, adjacent bonds internal to the hull, and adjacent bonds external to the hull. For the ninth generation, only the hull was obtained.\label{table:counter}}
\end{table}

\subsection{Counter-rotated model}
Data on the first eight generations of clusters, and the first nine generations of hulls, for the counter-rotated model are reported in~\autoref{table:counter}. Since there is only one species, the self-similar structure (as exhibited in~\autoref{fig:counter-cluster}) is much simpler than for the checkerboard model. The "length" $L$ of the cluster is the side length of the smallest square enclosing the cluster, whose sides are aligned with the horizontal and vertical axes in~\autoref{fig:lattice-counter}. Since all of the critical clusters are aligned with $\pm \pi/4$, i.e. along the $+$ or $-$ bonds of $\mathcal{L}$, the length along the major axis is $L\sqrt{2}$. As expected, $L(j)/L(j-1)$ converges to $2+\sqrt{3}$. $L(2),...,L(8)$ follow sequence A263942 in Ref.~\cite{oeis}. The other quantities do not follow any known sequences, but we empirically observe that $N_{out}(j)=N_{in}(j)+10$ for $m\geq 3$. Another quirk is that at $b=0$, the smallest clusters have mass 2, not 1. This means that for a given vertex, at least one of the incident edges has $b(x,y)\leq 0$. 

\section{Boundary conditions for the checkerboard model}\label{app:boundary}
In this appendix, we give details on the open and periodic boundary conditions used for finite size scaling in the checkerboard model. 

\subsection{Open boundary conditions}
\begin{figure}[t]
    \centering
    \includegraphics[width=\linewidth]{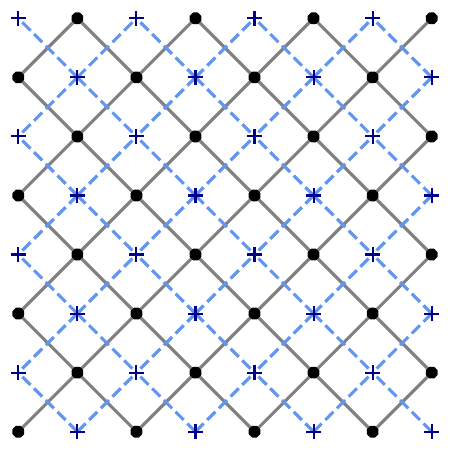}
    \caption{Lattice $\mathcal{L}$ (black, solid lines) and dual lattice (blue, dashed lines) for a system of nominal size $L_1=7$ and open boundary conditions, consisting of $L_1^2=49$ bonds and $(L_1+1)^2/2=32$ vertices.}
    \label{fig:obc}
\end{figure}

Since $\mathcal{L}$ and $\mathcal{L'}$ are not commensurate, if we do not distort $\mathcal{L'}$, we must use open boundary conditions, which we define to fulfill two conditions:  First, to take full advantage of the self-dual nature of the model in the bulk, the boundaries need to be fine-tuned so that the lattice is completely self-dual. This construction, which requires the nominal system size $L_1$ to be odd and results in a staggered lattice with $(L+1)/2$ vertices in each row, is shown in~\autoref{fig:obc}. The system should also have an aspect ratio of 1, so that the length of the top/bottom boundaries and left/right boundaries are all equal. This preserves the underlying symmetry under $\pi/2$ rotations in the absence of the anisotropic term.

\subsection{Periodic boundary conditions}
Using the discrete scale-invariant pattern of near-commensurate points on the Ford lattice, we also construct two sequences of rational approximants to $\vec{a}$ for use in periodic boundary conditions (PBCs) as described in the main text. To ensure that coordinates separated by integer multiples of $[L_1,L_2]$ and $[-L_2, L_1]$ are fully equivalent, these points must have equal parity with respect to both $\mathcal{L}$ and $\mathcal{L'}$. Equal parity on $\mathcal{L}$ means $L_1+L_2$ must be even, and ensures that the PBCs identify odd bonds with odd bonds and even with even. Equal parity on $\mathcal{L'}$ is needed to identify the same sublattices on the checkerboard, so letting $[L_1, L_2] = \vec{A} [l_1, l_2]$, $l_1+l_2$ must be even as well.

$L_1(j)$, $l_1(j)$ and $l_2(j)$ for each sequence of system sizes with PBCs all obey the same recursion relation:
\begin{align}\label{eq:recursive}
    x(j) &= 4 x(j-1) - x(j-2) \notag \\ \Rightarrow \lim_{j\rightarrow\infty} \frac{x(j)}{x(j-1)}&=2+\sqrt{3}
\end{align}

The "even" sequence of system sizes has $L_2^{e}=0$, so the periodic boundaries align with horizontal/vertical, with $L_1^{e}(1)=6, L_1^{e}(2)=22$~\cite{oeis}. The alternating sequence, denoted "odd," has $L_2^{o}=L_1^{o}$, so the periodic boundaries are rotated by $\pi/4$ with respect to horizontal/vertical, with
$L_1^{o}(1)=2,L_1^o(2)=8$. In either case, a system periodic under $[L_1,L_2]$ contains $|\mathcal{V}|=(L_1^2 + L_2^2)/2$ vertices, and up to $L_1^2 + L_2^2$ edges. The system size $L$ is then defined as $\sqrt{|\mathcal{V}|}$, although it is sometimes more illuminating to perform a scaling collapse with $L_1$.

$l_1(j)$ and $l_2(j)$ are related to these sequences of system sizes by:
\begin{subequations}
\begin{align}
l_1^{e}(0)&=1, l_1^{e}(1)=5 \qquad l_2^{e}(j)=-L_1^{e}(j)/2 \label{eq:even} \\
l_1^{o}(j)&=L_1^{e}(j+1)/2, \qquad l_2^{o}(j)=L_1^{e}(j)/2 \label{eq:odd}
\end{align}
\end{subequations}
Note that $l_1^{e}(j)/L_1^{e}(j)$ defines a series of lower principal convergents to $a_1=\sqrt{3}/2$ (see entry A001834 of Ref.~\cite{oeis}), while $l_2^{e}/L_1^{e}(j)=-a_2=-1/2$. Meanwhile $l_1^{o}(j)/L_1^{o}(j)$ defines a series of upper approximants to $a_1+a_2=(1+\sqrt{3})/2$, while $l_2^{o}(j)/L_1^{o}(j)$ defines a series of upper approximants to $a_1-a_2=(1-\sqrt{3})/2$. This means that for the odd parity sequence, $a_1$ and $a_2$ are adjusted slightly below their Ford values to make the periodic boundary conditions commensurate, whereas for the even parity sequence, $(a_1,a_2)$ are both slightly above their Ford values. Moreover, adjusting $\vec{a}$ to satisfy $[L_1, L_2]=\vec{A} [l_1, l_2]$ yields commensurate lattice vectors
\begin{equation}\label{eq:commensurate}
    \vec{d}_{\pm}=[(L_1\pm L_2)/2, (L_1 \mp L_2)/2],
\end{equation}
such that every A vertex on the checkerboard is partnered with a B vertex at a displacement of $\vec{d_\pm}$.

\section{Finite size scaling in the checkerboard model}\label{app:fss}

In the main text, we noted that the wrapping interval for PBCs gives rise to a spurious exponent $\nu=1/2$ when working in ensembles at fixed $n$. In this appendix, we expand on the origin of this exponent, detailing the scaling collapse of the wrapping probabilities and crossing probabilities for fixed $n$ as well as fixed $b$ ensembles.

\subsection{Distributions of b-scores}
The fraction of occupied bonds, $n=N/N_e$, where $N_e=2L^2$ is the number of available bonds and $N$ the number of occupied bonds, is a monotonic function of $b$:
\begin{equation}\label{eq:n(b)}
    n(b) = |\mathcal{E}(b)| = |\{(x,y): b(x,y) \leq b\}|
\end{equation}

Although $n(b)$ is monotonic, the presence of sharp steps in the distribution of $b(x,y)$ near $b=0$ on finite lattices leads to significantly different scaling functions $f(x)$ and $g(x)$. In particular, the fact that $n(b)$ is not a smooth function near $b=b_c$ leads to different exponents inferred from the wrapping probability (\autoref{eq:scaling-wrap}). To shed some light on this, we study the distribution of $b$ scores $b(x,y)$ vs. their rank.

\begin{figure}[t]
    \centering
    \subfloat[]{
    \includegraphics[width=0.5\linewidth]{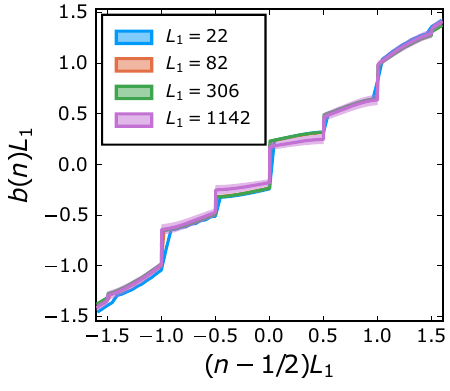}\label{fig:bsteps}
    }
    \subfloat[]{
    \includegraphics[width=0.5\linewidth]{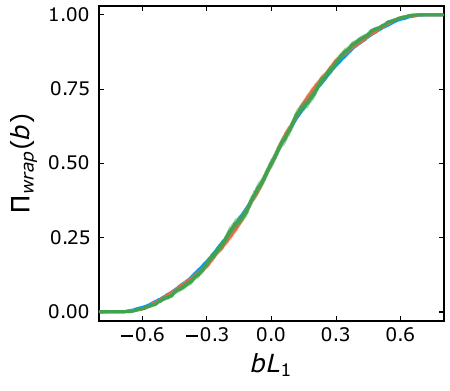}
    \label{fig:scaled-b}
    }
    \caption{(a) $b(n)$ averaged over samples for the even sequence of system sizes, $[L_1^e,0]$, near $n_c=1/2$ in the checkerboard Ford model. (b) Scaling collapse of the average wrapping probability, $\Pi_{wrap}(b)=\frac{1}{2} (\Pi^+(b) + \Pi^-(b))$, in fixed $b$ ensembles, implies a spurious exponent $\nu=1$. $L_1=1142$ not shown.}
\end{figure}

For PBCs, the distribution of $b$ scores is perfectly symmetric about 0, i.e. $b(n) = -b(1-n)$, due to the commensurate point in the middle of the lattice (\autoref{eq:commensurate}):
\begin{equation}
    b(x,y) = -b(\vec{d_\pm} + (x,y))
\end{equation}
Focusing on the vicinity of the wrapping threshold, there is a large step at $b=0$, with smaller steps nearby at discrete intervals. Different samples can have different widths and heights for the steps, but widths of the intervals between steps are concentrated around $\Delta n =1/2L_1$ as shown in~\autoref{fig:bsteps}.

For the even parity sequence of system sizes, $[L_1^e,0]$, the wrapping in one direction occurs at $b<0$ ($n<1/2$) when bond $e_-$ gets added and the wrapping in the other direction occurs at $b>0$ ($n=1/2$) when the next bond $e_+$ gets added. Thus the two bonds $e_-$, $e_+$ which mediate the wrapping are displaced by a commensurate lattice vector $\vec{d_\pm}$, and the wrapping events occur on either side of a large step in the $b$ scores. Since the steps in the $b$ scores, including the one centered at $b=0$, have an average height $\Delta b \propto 1/L$, finite size scaling of the wrapping interval in terms of $b$ would imply $\nu=1$. This can also be seen from the scaling collapse of $g(bL_1)$ (\autoref{fig:scaled-b}). While $\Pi_{wrap}(b)$ is a smooth function of $b$, this is solely due to the smooth distribution of $b$ scores within the central step, which has a tail all the way down to $\Delta b=0$. The deduced exponent from this scaling collapse is contrary to the exponent $\nu=1/2$ inferred from $\Delta N = 1$. Neither of these inferred exponents is the "true" critical exponent, however, since they only capture the finite size rounding of the step in the scaling function $f(x)$. 

It should be emphasized that the steps in $b(n)$ near $n=n_c$ are fundamentally a finite-size effect, not the consequence of adjusting $\mathcal{L'}$ to use PBCs. As $L\rightarrow \infty$, the height and width of the central step both go to 0 as $1/L$. In the limit of infinite system size, under the assumption that the vertices of $\mathbb{Z}^2$ are uniformly distributed within the enlarged unit cell of the checkerboard on $\mathcal{L'}$, which appears to hold when $\mathcal{L'}$ is the "maximally incommensurate" Ford lattice, $n(b)$ takes the somewhat unwieldy functional form:
\begin{widetext}
\begin{align}\label{eq:nb}
n(b) = \frac{1}{2} + \frac{b}{8 \beta_+ \beta_-} \bigg\{&- 2\gamma \left[2 b^2 \log\left(1+\gamma\sqrt{2}\right) + \beta_+ \beta_- \log\left(\frac{-1 + \abs{b}\beta_+ \sqrt{2} + \gamma \sqrt{2}}{\beta_+}\right)-\log\left(2\gamma + \sqrt{2}\right)\right] \notag \\
&+4 \sqrt{2} - 4\abs{b} \beta_+ - \gamma \sqrt{2} \bigg\}
\end{align}
\end{widetext}
where
\begin{align*}
    \beta_\pm = 1 \pm |b|\sqrt{2}, \qquad \gamma = \sqrt{1-b^2}
\end{align*}
Substituting $b\rightarrow b\sqrt{2}$ yields the expression for $n(b)$ in the counter-rotated model (for which $b$ ranges from $-1/\sqrt{2}$ to $1/\sqrt{2}$). From~\autoref{eq:nb}, $n(b)$ is continuous with a continuous first derivative at $b=0$. Therefore, one advantage of the nominally infinite methods used in the main text is that scaling as a function of $n$ and $b$ give consistent results for the critical exponent $\nu$, as in random percolation.

\subsection{Crossing probabilities}
\begin{figure}[t]
    \centering
    \includegraphics[width=\linewidth]{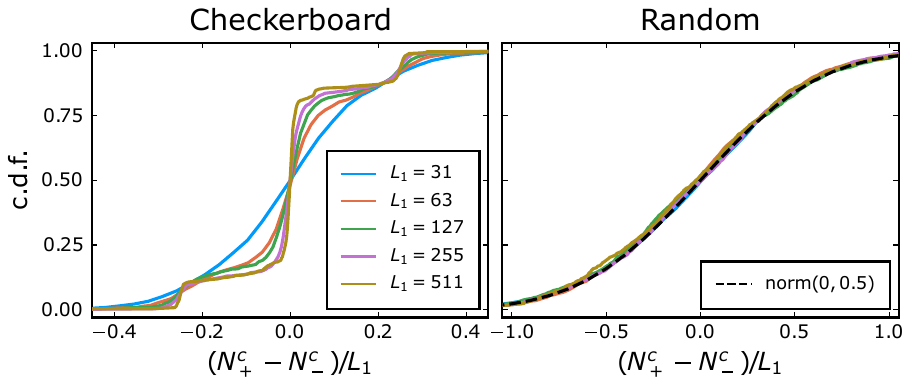}
    \caption{Cumulative distribution function of $(N_+^c-N_-^c)/L_1$ for quasiperiodic Ford model (left) vs. random bond percolation (right), with open boundary conditions. Black dashed curve is the cumulative normal distribution function with mean 0, standard deviation $1/2$.}
    \label{fig:obc-anisotropy}
\end{figure}
Turning to OBCs, while the typical spanning threshold in both directions is $n_c\approx 1/2$, the distribution of thresholds has secondary peaks at a distance of $\Delta n \propto 1/L$ above or below $1/2$. Thus, the scaling function $f$ for the average crossing probability, defined as $\frac{1}{2} (\Pi^+(n) + \Pi^-(n))$, consists of more than one step, unlike the wrapping probability. At the level of a single sample, while the majority of samples have a spanning interval $\Delta N = O(1)$, an extensive fraction instead have $\Delta N=O(L)$.

These large spanning intervals and concomitant secondary peaks in the threshold distribution originate primarily from a spontaneous inhomogeneity in the number of occupied $+$ and $-$ bonds at the threshold, denoted $N_\pm^c$. The cumulative distribution function of $(N_+^c-N_-^c)/L_1$ is plotted in~\autoref{fig:obc-anisotropy}, exhibiting a large central step along with secondary steps at $\pm 1/4$. Note that, aside from the fact that with OBCs the number of available (+) bonds is 1 more than the number of available ($-$) bonds (the opposite is true on the dual lattice), there is no microscopic bias toward either parity bond. For this reason, the secondary steps are of either sign, and we can think of the infinite system as containing patches with a spontaneous preference for either parity. This spontaneous imbalance is related to the static 1d inhomogeneity in the presence of anisotropy (\autoref{fig:obc-local}). The $O(L)$ surplus of $+$ or $-$ bonds can be thought of as producing an effective nonzero $n'$ of either sign in that patch of the system, which in turn implies the existence of columns or rows with anomalous local occupation rates. Indeed, in such patches, the large spanning interval arises when, after a cluster has already crossed in one direction, a strip of bonds all in the same row or column of the lattice and with the same parity are added consecutively before the cluster can cross in the other direction.

This is in contrast to PBCs, where $|N_+^c - N_-^c|\leq 1$ at each wrapping threshold, and the wrapping interval consists of adding just one bond of either parity rather than an entire row. It is also qualitatively different from what occurs for random bond percolation (right panel of~\autoref{fig:obc-anisotropy}). In that case, $N_+^c-N_-^c$ follows an approximately normal distribution, with variance $\sigma^2 = L_1^2/4$.


\begin{thebibliography}{49}%
\makeatletter
\providecommand \@ifxundefined [1]{%
 \@ifx{#1\undefined}
}%
\providecommand \@ifnum [1]{%
 \ifnum #1\expandafter \@firstoftwo
 \else \expandafter \@secondoftwo
 \fi
}%
\providecommand \@ifx [1]{%
 \ifx #1\expandafter \@firstoftwo
 \else \expandafter \@secondoftwo
 \fi
}%
\providecommand \natexlab [1]{#1}%
\providecommand \enquote  [1]{``#1''}%
\providecommand \bibnamefont  [1]{#1}%
\providecommand \bibfnamefont [1]{#1}%
\providecommand \citenamefont [1]{#1}%
\providecommand \href@noop [0]{\@secondoftwo}%
\providecommand \href [0]{\begingroup \@sanitize@url \@href}%
\providecommand \@href[1]{\@@startlink{#1}\@@href}%
\providecommand \@@href[1]{\endgroup#1\@@endlink}%
\providecommand \@sanitize@url [0]{\catcode `\\12\catcode `\$12\catcode
  `\&12\catcode `\#12\catcode `\^12\catcode `\_12\catcode `\%12\relax}%
\providecommand \@@startlink[1]{}%
\providecommand \@@endlink[0]{}%
\providecommand \url  [0]{\begingroup\@sanitize@url \@url }%
\providecommand \@url [1]{\endgroup\@href {#1}{\urlprefix }}%
\providecommand \urlprefix  [0]{URL }%
\providecommand \Eprint [0]{\href }%
\providecommand \doibase [0]{https://doi.org/}%
\providecommand \selectlanguage [0]{\@gobble}%
\providecommand \bibinfo  [0]{\@secondoftwo}%
\providecommand \bibfield  [0]{\@secondoftwo}%
\providecommand \translation [1]{[#1]}%
\providecommand \BibitemOpen [0]{}%
\providecommand \bibitemStop [0]{}%
\providecommand \bibitemNoStop [0]{.\EOS\space}%
\providecommand \EOS [0]{\spacefactor3000\relax}%
\providecommand \BibitemShut  [1]{\csname bibitem#1\endcsname}%
\let\auto@bib@innerbib\@empty
%</preamble>
\bibitem [{\citenamefont {{Kasteleyn}}\ and\ \citenamefont
  {{Fortuin}}(1969)}]{Kasteleyn1969}%
  \BibitemOpen
  \bibfield  {author} {\bibinfo {author} {\bibfnamefont {P.~W.}\ \bibnamefont
  {{Kasteleyn}}}\ and\ \bibinfo {author} {\bibfnamefont {C.~M.}\ \bibnamefont
  {{Fortuin}}},\ }\bibfield  {title} {\bibinfo {title} {{Phase Transitions in
  Lattice Systems with Random Local Properties}},\ }\href
  {https://ui.adsabs.harvard.edu/abs/1969PSJJS..26...11K} {\bibfield  {journal}
  {\bibinfo  {journal} {Physical Society of Japan Journal Supplement}\ }\textbf
  {\bibinfo {volume} {26}},\ \bibinfo {pages} {11} (\bibinfo {year}
  {1969})}\BibitemShut {NoStop}%
\bibitem [{\citenamefont {Wu}(1978)}]{Wu1978}%
  \BibitemOpen
  \bibfield  {author} {\bibinfo {author} {\bibfnamefont {F.~Y.}\ \bibnamefont
  {Wu}},\ }\bibfield  {title} {\bibinfo {title} {{Percolation and the Potts
  model}},\ }\href {https://doi.org/10.1007/BF01014303} {\bibfield  {journal}
  {\bibinfo  {journal} {Journal of Statistical Physics}\ }\textbf {\bibinfo
  {volume} {18}},\ \bibinfo {pages} {115} (\bibinfo {year} {1978})}\BibitemShut
  {NoStop}%
\bibitem [{\citenamefont {Cardy}(1992)}]{Cardy1992}%
  \BibitemOpen
  \bibfield  {author} {\bibinfo {author} {\bibfnamefont {J.~L.}\ \bibnamefont
  {Cardy}},\ }\bibfield  {title} {\bibinfo {title} {Critical percolation in
  finite geometries},\ }\href {https://doi.org/10.1088/0305-4470/25/4/009}
  {\bibfield  {journal} {\bibinfo  {journal} {Journal of Physics A:
  Mathematical and General}\ }\textbf {\bibinfo {volume} {25}},\ \bibinfo
  {pages} {L201} (\bibinfo {year} {1992})}\BibitemShut {NoStop}%
\bibitem [{\citenamefont {Langlands}\ \emph {et~al.}(1994)\citenamefont
  {Langlands}, \citenamefont {Pouliot},\ and\ \citenamefont
  {Saint-Aubin}}]{Langlands1994}%
  \BibitemOpen
  \bibfield  {author} {\bibinfo {author} {\bibfnamefont {R.}~\bibnamefont
  {Langlands}}, \bibinfo {author} {\bibfnamefont {P.}~\bibnamefont {Pouliot}},\
  and\ \bibinfo {author} {\bibfnamefont {Y.}~\bibnamefont {Saint-Aubin}},\
  }\bibfield  {title} {\bibinfo {title} {{Conformal invariance in
  two-dimensional percolation}},\ }\href
  {https://doi.org/10.1090/s0273-0979-1994-00456-2} {\bibfield  {journal}
  {\bibinfo  {journal} {Bulletin of the American Mathematical Society}\
  }\textbf {\bibinfo {volume} {30}},\ \bibinfo {pages} {1} (\bibinfo {year}
  {1994})}\BibitemShut {NoStop}%
\bibitem [{\citenamefont {Smirnov}(2001)}]{Smirnov2001}%
  \BibitemOpen
  \bibfield  {author} {\bibinfo {author} {\bibfnamefont {S.}~\bibnamefont
  {Smirnov}},\ }\bibfield  {title} {\bibinfo {title} {{Critical percolation in
  the plane: conformal invariance, Cardy's formula, scaling limits}},\ }\href
  {https://doi.org/10.1016/s0764-4442(01)01991-7} {\bibfield  {journal}
  {\bibinfo  {journal} {Comptes Rendus de l'Acad{\'{e}}mie des Sciences -
  Series I - Mathematics}\ }\textbf {\bibinfo {volume} {333}},\ \bibinfo
  {pages} {239} (\bibinfo {year} {2001})}\BibitemShut {NoStop}%
\bibitem [{\citenamefont {Duminil-Copin}\ \emph {et~al.}(2020)\citenamefont
  {Duminil-Copin}, \citenamefont {Kozlowski}, \citenamefont {Krachun},
  \citenamefont {Manolescu},\ and\ \citenamefont
  {Oulamara}}]{Duminil-Copin2020}%
  \BibitemOpen
  \bibfield  {author} {\bibinfo {author} {\bibfnamefont {H.}~\bibnamefont
  {Duminil-Copin}}, \bibinfo {author} {\bibfnamefont {K.~K.}\ \bibnamefont
  {Kozlowski}}, \bibinfo {author} {\bibfnamefont {D.}~\bibnamefont {Krachun}},
  \bibinfo {author} {\bibfnamefont {I.}~\bibnamefont {Manolescu}},\ and\
  \bibinfo {author} {\bibfnamefont {M.}~\bibnamefont {Oulamara}},\ }\href
  {http://arxiv.org/abs/2012.11672} {\bibinfo {title} {{Rotational invariance
  in critical planar lattice models}}} (\bibinfo {year} {2020}),\ \Eprint
  {https://arxiv.org/abs/2012.11672} {arXiv:2012.11672} \BibitemShut {NoStop}%
\bibitem [{\citenamefont {Kesten}(1980)}]{Kesten1980}%
  \BibitemOpen
  \bibfield  {author} {\bibinfo {author} {\bibfnamefont {H.}~\bibnamefont
  {Kesten}},\ }\bibfield  {title} {\bibinfo {title} {{The critical probability
  of bond percolation on the square lattice equals 1/2}},\ }\href
  {https://doi.org/10.1007/BF01197577} {\bibfield  {journal} {\bibinfo
  {journal} {Communications in Mathematical Physics}\ }\textbf {\bibinfo
  {volume} {74}},\ \bibinfo {pages} {41} (\bibinfo {year} {1980})}\BibitemShut
  {NoStop}%
\bibitem [{\citenamefont {Li}\ \emph {et~al.}(2018)\citenamefont {Li},
  \citenamefont {Chen},\ and\ \citenamefont {Fisher}}]{Li2018}%
  \BibitemOpen
  \bibfield  {author} {\bibinfo {author} {\bibfnamefont {Y.}~\bibnamefont
  {Li}}, \bibinfo {author} {\bibfnamefont {X.}~\bibnamefont {Chen}},\ and\
  \bibinfo {author} {\bibfnamefont {M.~P.~A.}\ \bibnamefont {Fisher}},\
  }\bibfield  {title} {\bibinfo {title} {{Quantum Zeno effect and the many-body
  entanglement transition}},\ }\href
  {https://doi.org/10.1103/PhysRevB.98.205136} {\bibfield  {journal} {\bibinfo
  {journal} {Physical Review B}\ }\textbf {\bibinfo {volume} {98}},\ \bibinfo
  {pages} {205136} (\bibinfo {year} {2018})}\BibitemShut {NoStop}%
\bibitem [{\citenamefont {Skinner}\ \emph {et~al.}(2019)\citenamefont
  {Skinner}, \citenamefont {Ruhman},\ and\ \citenamefont
  {Nahum}}]{Skinner2018}%
  \BibitemOpen
  \bibfield  {author} {\bibinfo {author} {\bibfnamefont {B.}~\bibnamefont
  {Skinner}}, \bibinfo {author} {\bibfnamefont {J.}~\bibnamefont {Ruhman}},\
  and\ \bibinfo {author} {\bibfnamefont {A.}~\bibnamefont {Nahum}},\ }\bibfield
   {title} {\bibinfo {title} {{Measurement-Induced Phase Transitions in the
  Dynamics of Entanglement}},\ }\href
  {https://doi.org/10.1103/PhysRevX.9.031009} {\bibfield  {journal} {\bibinfo
  {journal} {Physical Review X}\ }\textbf {\bibinfo {volume} {9}},\ \bibinfo
  {pages} {031009} (\bibinfo {year} {2019})}\BibitemShut {NoStop}%
\bibitem [{\citenamefont {Li}\ \emph {et~al.}(2019)\citenamefont {Li},
  \citenamefont {Chen},\ and\ \citenamefont {Fisher}}]{Li2019}%
  \BibitemOpen
  \bibfield  {author} {\bibinfo {author} {\bibfnamefont {Y.}~\bibnamefont
  {Li}}, \bibinfo {author} {\bibfnamefont {X.}~\bibnamefont {Chen}},\ and\
  \bibinfo {author} {\bibfnamefont {M.~P.~A.}\ \bibnamefont {Fisher}},\
  }\bibfield  {title} {\bibinfo {title} {{Measurement-driven entanglement
  transition in hybrid quantum circuits}},\ }\href
  {https://doi.org/10.1103/PhysRevB.100.134306} {\bibfield  {journal} {\bibinfo
   {journal} {Physical Review B}\ }\textbf {\bibinfo {volume} {100}},\ \bibinfo
  {pages} {134306} (\bibinfo {year} {2019})}\BibitemShut {NoStop}%
\bibitem [{\citenamefont {Chan}\ \emph {et~al.}(2019)\citenamefont {Chan},
  \citenamefont {Nandkishore}, \citenamefont {Pretko},\ and\ \citenamefont
  {Smith}}]{Chan2019}%
  \BibitemOpen
  \bibfield  {author} {\bibinfo {author} {\bibfnamefont {A.}~\bibnamefont
  {Chan}}, \bibinfo {author} {\bibfnamefont {R.~M.}\ \bibnamefont
  {Nandkishore}}, \bibinfo {author} {\bibfnamefont {M.}~\bibnamefont
  {Pretko}},\ and\ \bibinfo {author} {\bibfnamefont {G.}~\bibnamefont
  {Smith}},\ }\bibfield  {title} {\bibinfo {title} {{Unitary-projective
  entanglement dynamics}},\ }\href {https://doi.org/10.1103/PhysRevB.99.224307}
  {\bibfield  {journal} {\bibinfo  {journal} {Physical Review B}\ }\textbf
  {\bibinfo {volume} {99}},\ \bibinfo {pages} {224307} (\bibinfo {year}
  {2019})}\BibitemShut {NoStop}%
\bibitem [{\citenamefont {Gullans}\ and\ \citenamefont
  {Huse}(2020{\natexlab{a}})}]{Gullans2020}%
  \BibitemOpen
  \bibfield  {author} {\bibinfo {author} {\bibfnamefont {M.~J.}\ \bibnamefont
  {Gullans}}\ and\ \bibinfo {author} {\bibfnamefont {D.~A.}\ \bibnamefont
  {Huse}},\ }\bibfield  {title} {\bibinfo {title} {{Dynamical Purification
  Phase Transition Induced by Quantum Measurements}},\ }\href
  {https://doi.org/10.1103/PhysRevX.10.041020} {\bibfield  {journal} {\bibinfo
  {journal} {Physical Review X}\ }\textbf {\bibinfo {volume} {10}},\ \bibinfo
  {pages} {041020} (\bibinfo {year} {2020}{\natexlab{a}})}\BibitemShut
  {NoStop}%
\bibitem [{\citenamefont {Jian}\ \emph {et~al.}(2020)\citenamefont {Jian},
  \citenamefont {You}, \citenamefont {Vasseur},\ and\ \citenamefont
  {Ludwig}}]{Jian2020}%
  \BibitemOpen
  \bibfield  {author} {\bibinfo {author} {\bibfnamefont {C.-M.}\ \bibnamefont
  {Jian}}, \bibinfo {author} {\bibfnamefont {Y.-Z.}\ \bibnamefont {You}},
  \bibinfo {author} {\bibfnamefont {R.}~\bibnamefont {Vasseur}},\ and\ \bibinfo
  {author} {\bibfnamefont {A.~W.~W.}\ \bibnamefont {Ludwig}},\ }\bibfield
  {title} {\bibinfo {title} {{Measurement-induced criticality in random quantum
  circuits}},\ }\href {https://doi.org/10.1103/PhysRevB.101.104302} {\bibfield
  {journal} {\bibinfo  {journal} {Physical Review B}\ }\textbf {\bibinfo
  {volume} {101}},\ \bibinfo {pages} {104302} (\bibinfo {year}
  {2020})}\BibitemShut {NoStop}%
\bibitem [{\citenamefont {Bao}\ \emph {et~al.}(2020)\citenamefont {Bao},
  \citenamefont {Choi},\ and\ \citenamefont {Altman}}]{Bao2020}%
  \BibitemOpen
  \bibfield  {author} {\bibinfo {author} {\bibfnamefont {Y.}~\bibnamefont
  {Bao}}, \bibinfo {author} {\bibfnamefont {S.}~\bibnamefont {Choi}},\ and\
  \bibinfo {author} {\bibfnamefont {E.}~\bibnamefont {Altman}},\ }\bibfield
  {title} {\bibinfo {title} {{Theory of the phase transition in random unitary
  circuits with measurements}},\ }\href
  {https://doi.org/10.1103/PhysRevB.101.104301} {\bibfield  {journal} {\bibinfo
   {journal} {Physical Review B}\ }\textbf {\bibinfo {volume} {101}},\ \bibinfo
  {pages} {104301} (\bibinfo {year} {2020})}\BibitemShut {NoStop}%
\bibitem [{\citenamefont {Gullans}\ and\ \citenamefont
  {Huse}(2020{\natexlab{b}})}]{Gullans2019}%
  \BibitemOpen
  \bibfield  {author} {\bibinfo {author} {\bibfnamefont {M.~J.}\ \bibnamefont
  {Gullans}}\ and\ \bibinfo {author} {\bibfnamefont {D.~A.}\ \bibnamefont
  {Huse}},\ }\bibfield  {title} {\bibinfo {title} {{Scalable Probes of
  Measurement-Induced Criticality}},\ }\href
  {https://doi.org/10.1103/PhysRevLett.125.070606} {\bibfield  {journal}
  {\bibinfo  {journal} {Physical Review Letters}\ }\textbf {\bibinfo {volume}
  {125}},\ \bibinfo {pages} {070606} (\bibinfo {year}
  {2020}{\natexlab{b}})}\BibitemShut {NoStop}%
\bibitem [{\citenamefont {Chernikov}\ and\ \citenamefont
  {Rogalsky}(1994)}]{Chernikov1994}%
  \BibitemOpen
  \bibfield  {author} {\bibinfo {author} {\bibfnamefont {A.~A.}\ \bibnamefont
  {Chernikov}}\ and\ \bibinfo {author} {\bibfnamefont {A.~V.}\ \bibnamefont
  {Rogalsky}},\ }\bibfield  {title} {\bibinfo {title} {{Stochastic webs and
  continuum percolation in quasiperiodic media}},\ }\href
  {https://doi.org/10.1063/1.166055} {\bibfield  {journal} {\bibinfo  {journal}
  {Chaos}\ }\textbf {\bibinfo {volume} {4}},\ \bibinfo {pages} {35} (\bibinfo
  {year} {1994})}\BibitemShut {NoStop}%
\bibitem [{\citenamefont {Ford}(1925)}]{Ford1925}%
  \BibitemOpen
  \bibfield  {author} {\bibinfo {author} {\bibfnamefont {L.~R.}\ \bibnamefont
  {Ford}},\ }\bibfield  {title} {\bibinfo {title} {{On the Closeness of
  Approach of Complex Rational Fractions to a Complex Irrational Number}},\
  }\href {https://doi.org/10.2307/1989059} {\bibfield  {journal} {\bibinfo
  {journal} {Transactions of the American Mathematical Society}\ }\textbf
  {\bibinfo {volume} {27}},\ \bibinfo {pages} {146} (\bibinfo {year}
  {1925})}\BibitemShut {NoStop}%
\bibitem [{\citenamefont {Perron}(1930)}]{Perron1930}%
  \BibitemOpen
  \bibfield  {author} {\bibinfo {author} {\bibfnamefont {O.}~\bibnamefont
  {Perron}},\ }\bibfield  {title} {\bibinfo {title} {{{\"{U}}ber die
  Approximation einer komplexen Zahl durch Zahlen des
  K{\"{o}}rpers$\mathfrak{K}(i)$}},\ }\href
  {https://doi.org/10.1007/BF01455709} {\bibfield  {journal} {\bibinfo
  {journal} {Mathematische Annalen}\ }\textbf {\bibinfo {volume} {103}},\
  \bibinfo {pages} {533} (\bibinfo {year} {1930})}\BibitemShut {NoStop}%
\bibitem [{\citenamefont {Schmidt}(1967)}]{Schmidt1967}%
  \BibitemOpen
  \bibfield  {author} {\bibinfo {author} {\bibfnamefont {A.~L.}\ \bibnamefont
  {Schmidt}},\ }\bibfield  {title} {\bibinfo {title} {{Farey Triangles and
  Farey Quadrangels in the Complex Plane.}},\ }\href
  {https://doi.org/10.7146/math.scand.a-10863} {\bibfield  {journal} {\bibinfo
  {journal} {MATHEMATICA SCANDINAVICA}\ }\textbf {\bibinfo {volume} {21}},\
  \bibinfo {pages} {241} (\bibinfo {year} {1967})}\BibitemShut {NoStop}%
\bibitem [{oei(2022)}]{oeis}%
  \BibitemOpen
  \href {http://oeis.org} {\bibinfo {title} {{T}he on-line encyclopedia of
  integer sequences}} (\bibinfo {year} {2022})\BibitemShut {NoStop}%
\bibitem [{\citenamefont {Leath}(1976)}]{Leath1976}%
  \BibitemOpen
  \bibfield  {author} {\bibinfo {author} {\bibfnamefont {P.~L.}\ \bibnamefont
  {Leath}},\ }\bibfield  {title} {\bibinfo {title} {{Cluster size and boundary
  distribution near percolation threshold}},\ }\href
  {https://doi.org/10.1103/PhysRevB.14.5046} {\bibfield  {journal} {\bibinfo
  {journal} {Physical Review B}\ }\textbf {\bibinfo {volume} {14}},\ \bibinfo
  {pages} {5046} (\bibinfo {year} {1976})}\BibitemShut {NoStop}%
\bibitem [{\citenamefont {Vyssotsky}\ \emph {et~al.}(1961)\citenamefont
  {Vyssotsky}, \citenamefont {Gordon}, \citenamefont {Frisch},\ and\
  \citenamefont {Hammersley}}]{Vyssotsky1961}%
  \BibitemOpen
  \bibfield  {author} {\bibinfo {author} {\bibfnamefont {V.~A.}\ \bibnamefont
  {Vyssotsky}}, \bibinfo {author} {\bibfnamefont {S.~B.}\ \bibnamefont
  {Gordon}}, \bibinfo {author} {\bibfnamefont {H.~L.}\ \bibnamefont {Frisch}},\
  and\ \bibinfo {author} {\bibfnamefont {J.~M.}\ \bibnamefont {Hammersley}},\
  }\bibfield  {title} {\bibinfo {title} {{Critical Percolation Probabilities
  (Bond Problem)}},\ }\href {https://doi.org/10.1103/PhysRev.123.1566}
  {\bibfield  {journal} {\bibinfo  {journal} {Physical Review}\ }\textbf
  {\bibinfo {volume} {123}},\ \bibinfo {pages} {1566} (\bibinfo {year}
  {1961})}\BibitemShut {NoStop}%
\bibitem [{\citenamefont {Grassberger}(1983)}]{Grassberger1983}%
  \BibitemOpen
  \bibfield  {author} {\bibinfo {author} {\bibfnamefont {P.}~\bibnamefont
  {Grassberger}},\ }\bibfield  {title} {\bibinfo {title} {{On the critical
  behavior of the general epidemic process and dynamical percolation}},\ }\href
  {https://doi.org/10.1016/0025-5564(82)90036-0} {\bibfield  {journal}
  {\bibinfo  {journal} {Mathematical Biosciences}\ }\textbf {\bibinfo {volume}
  {63}},\ \bibinfo {pages} {157} (\bibinfo {year} {1983})}\BibitemShut
  {NoStop}%
\bibitem [{\citenamefont {Ziff}(2021)}]{Ziff2021}%
  \BibitemOpen
  \bibfield  {author} {\bibinfo {author} {\bibfnamefont {R.~M.}\ \bibnamefont
  {Ziff}},\ }\bibfield  {title} {\bibinfo {title} {{Efficient Simulation of
  Percolation Lattices}},\ }in\ \href
  {https://doi.org/10.1007/978-1-0716-1457-0} {\emph {\bibinfo {booktitle}
  {Complex Media and Percolation Theory}}},\ \bibinfo {editor} {edited by\
  \bibinfo {editor} {\bibfnamefont {M.}~\bibnamefont {Sahimi}}\ and\ \bibinfo
  {editor} {\bibfnamefont {A.~G.}\ \bibnamefont {Hunt}}}\ (\bibinfo
  {publisher} {Springer US},\ \bibinfo {address} {New York, NY},\ \bibinfo
  {year} {2021})\ pp.\ \bibinfo {pages} {25--47}\BibitemShut {NoStop}%
\bibitem [{\citenamefont {Newman}\ and\ \citenamefont
  {Ziff}(2000)}]{Newman2000}%
  \BibitemOpen
  \bibfield  {author} {\bibinfo {author} {\bibfnamefont {M.~E.~J.}\
  \bibnamefont {Newman}}\ and\ \bibinfo {author} {\bibfnamefont {R.~M.}\
  \bibnamefont {Ziff}},\ }\bibfield  {title} {\bibinfo {title} {{Efficient
  Monte Carlo Algorithm and High-Precision Results for Percolation}},\ }\href
  {https://doi.org/10.1103/PhysRevLett.85.4104} {\bibfield  {journal} {\bibinfo
   {journal} {Physical Review Letters}\ }\textbf {\bibinfo {volume} {85}},\
  \bibinfo {pages} {4104} (\bibinfo {year} {2000})}\BibitemShut {NoStop}%
\bibitem [{\citenamefont {Newman}\ and\ \citenamefont
  {Ziff}(2001)}]{Newman2001}%
  \BibitemOpen
  \bibfield  {author} {\bibinfo {author} {\bibfnamefont {M.~E.~J.}\
  \bibnamefont {Newman}}\ and\ \bibinfo {author} {\bibfnamefont {R.~M.}\
  \bibnamefont {Ziff}},\ }\bibfield  {title} {\bibinfo {title} {{Fast Monte
  Carlo algorithm for site or bond percolation}},\ }\href
  {https://doi.org/10.1103/PhysRevE.64.016706} {\bibfield  {journal} {\bibinfo
  {journal} {Physical Review E}\ }\textbf {\bibinfo {volume} {64}},\ \bibinfo
  {pages} {016706} (\bibinfo {year} {2001})}\BibitemShut {NoStop}%
\bibitem [{\citenamefont {Christensen}(2002)}]{Christensen2002}%
  \BibitemOpen
  \bibfield  {author} {\bibinfo {author} {\bibfnamefont {K.}~\bibnamefont
  {Christensen}},\ }\href
  {http://www.mit.edu/$\sim$levitov/8.334/notes/percol_notes.pdf} {\bibinfo
  {title} {{Percolation theory}}} (\bibinfo {year} {2002})\BibitemShut
  {NoStop}%
\bibitem [{\citenamefont {Voss}(1984)}]{Voss1984}%
  \BibitemOpen
  \bibfield  {author} {\bibinfo {author} {\bibfnamefont {R.~F.}\ \bibnamefont
  {Voss}},\ }\bibfield  {title} {\bibinfo {title} {{The fractal dimension of
  percolation cluster hulls}},\ }\bibfield  {journal} {\bibinfo  {journal}
  {Journal of Physics A: General Physics}\ }\textbf {\bibinfo {volume} {17}},\
  \href {https://doi.org/10.1088/0305-4470/17/7/001}
  {10.1088/0305-4470/17/7/001} (\bibinfo {year} {1984})\BibitemShut {NoStop}%
\bibitem [{\citenamefont {Ziff}\ \emph {et~al.}(1984)\citenamefont {Ziff},
  \citenamefont {Cummings},\ and\ \citenamefont {Stells}}]{Ziff1984}%
  \BibitemOpen
  \bibfield  {author} {\bibinfo {author} {\bibfnamefont {R.~M.}\ \bibnamefont
  {Ziff}}, \bibinfo {author} {\bibfnamefont {P.~T.}\ \bibnamefont {Cummings}},\
  and\ \bibinfo {author} {\bibfnamefont {G.}~\bibnamefont {Stells}},\
  }\bibfield  {title} {\bibinfo {title} {{Generation of percolation cluster
  perimeters by a random walk}},\ }\href
  {https://doi.org/10.1088/0305-4470/17/15/018} {\bibfield  {journal} {\bibinfo
   {journal} {Journal of Physics A: Mathematical and General}\ }\textbf
  {\bibinfo {volume} {17}},\ \bibinfo {pages} {3009} (\bibinfo {year}
  {1984})}\BibitemShut {NoStop}%
\bibitem [{\citenamefont {Weinrib}\ and\ \citenamefont
  {Trugman}(1985)}]{Weinrib1985}%
  \BibitemOpen
  \bibfield  {author} {\bibinfo {author} {\bibfnamefont {A.}~\bibnamefont
  {Weinrib}}\ and\ \bibinfo {author} {\bibfnamefont {S.~A.}\ \bibnamefont
  {Trugman}},\ }\bibfield  {title} {\bibinfo {title} {{A new kinetic walk and
  percolation perimeters}},\ }\href {https://doi.org/10.1103/PhysRevB.31.2993}
  {\bibfield  {journal} {\bibinfo  {journal} {Physical Review B}\ }\textbf
  {\bibinfo {volume} {31}},\ \bibinfo {pages} {2993} (\bibinfo {year}
  {1985})}\BibitemShut {NoStop}%
\bibitem [{\citenamefont {Grassberger}(1986)}]{Grassberger1986}%
  \BibitemOpen
  \bibfield  {author} {\bibinfo {author} {\bibfnamefont {P.}~\bibnamefont
  {Grassberger}},\ }\bibfield  {title} {\bibinfo {title} {{On the hull of
  two-dimensional percolation clusters}},\ }\href
  {https://doi.org/10.1088/0305-4470/19/13/032} {\bibfield  {journal} {\bibinfo
   {journal} {Journal of Physics A: Mathematical and General}\ }\textbf
  {\bibinfo {volume} {19}},\ \bibinfo {pages} {2675} (\bibinfo {year}
  {1986})}\BibitemShut {NoStop}%
\bibitem [{\citenamefont {Machta}\ \emph {et~al.}(1996)\citenamefont {Machta},
  \citenamefont {Choi}, \citenamefont {Lucke}, \citenamefont {Schweizer},\ and\
  \citenamefont {Chayes}}]{Machta1996}%
  \BibitemOpen
  \bibfield  {author} {\bibinfo {author} {\bibfnamefont {J.}~\bibnamefont
  {Machta}}, \bibinfo {author} {\bibfnamefont {Y.~S.}\ \bibnamefont {Choi}},
  \bibinfo {author} {\bibfnamefont {A.}~\bibnamefont {Lucke}}, \bibinfo
  {author} {\bibfnamefont {T.}~\bibnamefont {Schweizer}},\ and\ \bibinfo
  {author} {\bibfnamefont {L.~M.}\ \bibnamefont {Chayes}},\ }\bibfield  {title}
  {\bibinfo {title} {{Invaded cluster algorithm for Potts models}},\ }\href
  {https://doi.org/10.1103/PhysRevE.54.1332} {\bibfield  {journal} {\bibinfo
  {journal} {Physical Review E - Statistical Physics, Plasmas, Fluids, and
  Related Interdisciplinary Topics}\ }\textbf {\bibinfo {volume} {54}},\
  \bibinfo {pages} {1332} (\bibinfo {year} {1996})}\BibitemShut {NoStop}%
\bibitem [{\citenamefont {{\v{S}}kvor}\ \emph {et~al.}(2007)\citenamefont
  {{\v{S}}kvor}, \citenamefont {Nezbeda}, \citenamefont {Brovchenko},\ and\
  \citenamefont {Oleinikova}}]{Skvor2007}%
  \BibitemOpen
  \bibfield  {author} {\bibinfo {author} {\bibfnamefont {J.}~\bibnamefont
  {{\v{S}}kvor}}, \bibinfo {author} {\bibfnamefont {I.}~\bibnamefont
  {Nezbeda}}, \bibinfo {author} {\bibfnamefont {I.}~\bibnamefont
  {Brovchenko}},\ and\ \bibinfo {author} {\bibfnamefont {A.}~\bibnamefont
  {Oleinikova}},\ }\bibfield  {title} {\bibinfo {title} {{Percolation
  Transition in Fluids: Scaling Behavior of the Spanning Probability
  Functions}},\ }\href {https://doi.org/10.1103/PhysRevLett.99.127801}
  {\bibfield  {journal} {\bibinfo  {journal} {Physical Review Letters}\
  }\textbf {\bibinfo {volume} {99}},\ \bibinfo {pages} {127801} (\bibinfo
  {year} {2007})}\BibitemShut {NoStop}%
\bibitem [{\citenamefont {Mertens}\ and\ \citenamefont
  {Moore}(2012)}]{Mertens2012}%
  \BibitemOpen
  \bibfield  {author} {\bibinfo {author} {\bibfnamefont {S.}~\bibnamefont
  {Mertens}}\ and\ \bibinfo {author} {\bibfnamefont {C.}~\bibnamefont
  {Moore}},\ }\bibfield  {title} {\bibinfo {title} {{Continuum percolation
  thresholds in two dimensions}},\ }\href
  {https://doi.org/10.1103/PhysRevE.86.061109} {\bibfield  {journal} {\bibinfo
  {journal} {Physical Review E}\ }\textbf {\bibinfo {volume} {86}},\ \bibinfo
  {pages} {061109} (\bibinfo {year} {2012})}\BibitemShut {NoStop}%
\bibitem [{\citenamefont {Stauffer}(1979)}]{Stauffer1979}%
  \BibitemOpen
  \bibfield  {author} {\bibinfo {author} {\bibfnamefont {D.}~\bibnamefont
  {Stauffer}},\ }\bibfield  {title} {\bibinfo {title} {{Scaling theory of
  percolation clusters}},\ }\href
  {https://doi.org/10.1016/0370-1573(79)90060-7} {\bibfield  {journal}
  {\bibinfo  {journal} {Physics Reports}\ }\textbf {\bibinfo {volume} {54}},\
  \bibinfo {pages} {1} (\bibinfo {year} {1979})}\BibitemShut {NoStop}%
\bibitem [{\citenamefont {Saleur}\ and\ \citenamefont
  {Duplantier}(1987)}]{Saleur1987}%
  \BibitemOpen
  \bibfield  {author} {\bibinfo {author} {\bibfnamefont {H.}~\bibnamefont
  {Saleur}}\ and\ \bibinfo {author} {\bibfnamefont {B.}~\bibnamefont
  {Duplantier}},\ }\bibfield  {title} {\bibinfo {title} {{Exact determination
  of the percolation hull exponent in two dimensions}},\ }\href
  {https://doi.org/10.1103/PhysRevLett.58.2325} {\bibfield  {journal} {\bibinfo
   {journal} {Physical Review Letters}\ }\textbf {\bibinfo {volume} {58}},\
  \bibinfo {pages} {2325} (\bibinfo {year} {1987})}\BibitemShut {NoStop}%
\bibitem [{\citenamefont {Lorenz}\ and\ \citenamefont
  {Ziff}(1998)}]{Lorenz1998}%
  \BibitemOpen
  \bibfield  {author} {\bibinfo {author} {\bibfnamefont {C.~D.}\ \bibnamefont
  {Lorenz}}\ and\ \bibinfo {author} {\bibfnamefont {R.~M.}\ \bibnamefont
  {Ziff}},\ }\bibfield  {title} {\bibinfo {title} {{Precise determination of
  the bond percolation thresholds and finite-size scaling corrections for the
  sc, fcc, and bcc lattices}},\ }\href
  {https://doi.org/10.1103/PhysRevE.57.230} {\bibfield  {journal} {\bibinfo
  {journal} {Physical Review E - Statistical Physics, Plasmas, Fluids, and
  Related Interdisciplinary Topics}\ }\textbf {\bibinfo {volume} {57}},\
  \bibinfo {pages} {230} (\bibinfo {year} {1998})}\BibitemShut {NoStop}%
\bibitem [{\citenamefont {Lorenz}\ \emph {et~al.}(2000)\citenamefont {Lorenz},
  \citenamefont {May},\ and\ \citenamefont {Ziff}}]{Lorenz2000}%
  \BibitemOpen
  \bibfield  {author} {\bibinfo {author} {\bibfnamefont {C.~D.}\ \bibnamefont
  {Lorenz}}, \bibinfo {author} {\bibfnamefont {R.}~\bibnamefont {May}},\ and\
  \bibinfo {author} {\bibfnamefont {R.~M.}\ \bibnamefont {Ziff}},\ }\bibfield
  {title} {\bibinfo {title} {{Similarity of percolation thresholds on the HCP
  and FCC lattices}},\ }\href {https://doi.org/10.1023/A:1018648130343}
  {\bibfield  {journal} {\bibinfo  {journal} {Journal of Statistical Physics}\
  }\textbf {\bibinfo {volume} {98}},\ \bibinfo {pages} {961} (\bibinfo {year}
  {2000})}\BibitemShut {NoStop}%
\bibitem [{\citenamefont {Harris}(1974)}]{Harris1974}%
  \BibitemOpen
  \bibfield  {author} {\bibinfo {author} {\bibfnamefont {A.~B.}\ \bibnamefont
  {Harris}},\ }\bibfield  {title} {\bibinfo {title} {{Effect of random defects
  on the critical behaviour of Ising models}},\ }\href
  {https://doi.org/10.1088/0022-3719/7/9/009} {\bibfield  {journal} {\bibinfo
  {journal} {Journal of Physics C: Solid State Physics}\ }\textbf {\bibinfo
  {volume} {7}},\ \bibinfo {pages} {1671} (\bibinfo {year} {1974})}\BibitemShut
  {NoStop}%
\bibitem [{\citenamefont {Ziff}(2010)}]{Ziff2010}%
  \BibitemOpen
  \bibfield  {author} {\bibinfo {author} {\bibfnamefont {R.~M.}\ \bibnamefont
  {Ziff}},\ }\bibfield  {title} {\bibinfo {title} {{Scaling behavior of
  explosive percolation on the square lattice}},\ }\href
  {https://doi.org/10.1103/PhysRevE.82.051105} {\bibfield  {journal} {\bibinfo
  {journal} {Physical Review E}\ }\textbf {\bibinfo {volume} {82}},\ \bibinfo
  {pages} {051105} (\bibinfo {year} {2010})}\BibitemShut {NoStop}%
\bibitem [{\citenamefont {Sykes}\ and\ \citenamefont
  {Essam}(1963)}]{Sykes1963}%
  \BibitemOpen
  \bibfield  {author} {\bibinfo {author} {\bibfnamefont {M.~F.}\ \bibnamefont
  {Sykes}}\ and\ \bibinfo {author} {\bibfnamefont {J.~W.}\ \bibnamefont
  {Essam}},\ }\bibfield  {title} {\bibinfo {title} {{Some Exact Critical
  Percolation Probabilities for Bond and Site Problems in Two Dimensions}},\
  }\href {https://doi.org/10.1103/PhysRevLett.10.3} {\bibfield  {journal}
  {\bibinfo  {journal} {Physical Review Letters}\ }\textbf {\bibinfo {volume}
  {10}},\ \bibinfo {pages} {3} (\bibinfo {year} {1963})}\BibitemShut {NoStop}%
\bibitem [{\citenamefont {Sykes}\ and\ \citenamefont
  {Essam}(1964)}]{Sykes1964}%
  \BibitemOpen
  \bibfield  {author} {\bibinfo {author} {\bibfnamefont {M.~F.}\ \bibnamefont
  {Sykes}}\ and\ \bibinfo {author} {\bibfnamefont {J.~W.}\ \bibnamefont
  {Essam}},\ }\bibfield  {title} {\bibinfo {title} {{Exact Critical Percolation
  Probabilities for Site and Bond Problems in Two Dimensions}},\ }\href
  {https://doi.org/10.1063/1.1704215} {\bibfield  {journal} {\bibinfo
  {journal} {Journal of Mathematical Physics}\ }\textbf {\bibinfo {volume}
  {5}},\ \bibinfo {pages} {1117} (\bibinfo {year} {1964})}\BibitemShut
  {NoStop}%
\bibitem [{\citenamefont {Temperley}\ and\ \citenamefont
  {Lieb}(1971)}]{Temperley1971}%
  \BibitemOpen
  \bibfield  {author} {\bibinfo {author} {\bibfnamefont {H.~N.~V.}\
  \bibnamefont {Temperley}}\ and\ \bibinfo {author} {\bibfnamefont {E.~H.}\
  \bibnamefont {Lieb}},\ }\bibfield  {title} {\bibinfo {title} {{Relations
  between the ‘percolation' and ‘colouring' problem and other
  graph-theoretical problems associated with regular planar lattices: some
  exact results for the ‘percolation' problem}},\ }\href
  {https://doi.org/10.1098/rspa.1971.0067} {\bibfield  {journal} {\bibinfo
  {journal} {Proceedings of the Royal Society of London. A. Mathematical and
  Physical Sciences}\ }\textbf {\bibinfo {volume} {322}},\ \bibinfo {pages}
  {251} (\bibinfo {year} {1971})}\BibitemShut {NoStop}%
\bibitem [{\citenamefont {Redner}\ and\ \citenamefont
  {Stanley}(1979)}]{Redner1979}%
  \BibitemOpen
  \bibfield  {author} {\bibinfo {author} {\bibfnamefont {S.}~\bibnamefont
  {Redner}}\ and\ \bibinfo {author} {\bibfnamefont {H.~E.}\ \bibnamefont
  {Stanley}},\ }\bibfield  {title} {\bibinfo {title} {{Anisotropic bond
  percolation}},\ }\href {https://doi.org/10.1088/0305-4470/12/8/021}
  {\bibfield  {journal} {\bibinfo  {journal} {Journal of Physics A:
  Mathematical and General}\ }\textbf {\bibinfo {volume} {12}},\ \bibinfo
  {pages} {1267} (\bibinfo {year} {1979})}\BibitemShut {NoStop}%
\bibitem [{\citenamefont {Devakul}\ and\ \citenamefont
  {Huse}(2017)}]{Devakul2017}%
  \BibitemOpen
  \bibfield  {author} {\bibinfo {author} {\bibfnamefont {T.}~\bibnamefont
  {Devakul}}\ and\ \bibinfo {author} {\bibfnamefont {D.~A.}\ \bibnamefont
  {Huse}},\ }\bibfield  {title} {\bibinfo {title} {{Anderson localization
  transitions with and without random potentials}},\ }\href
  {https://doi.org/10.1103/PhysRevB.96.214201} {\bibfield  {journal} {\bibinfo
  {journal} {Physical Review B}\ }\textbf {\bibinfo {volume} {96}},\ \bibinfo
  {pages} {214201} (\bibinfo {year} {2017})}\BibitemShut {NoStop}%
\bibitem [{\citenamefont {Sommers}\ \emph {et~al.}(2022)\citenamefont
  {Sommers}, \citenamefont {Huse},\ and\ \citenamefont
  {Gullans}}]{Sommers2022b}%
  \BibitemOpen
  \bibfield  {author} {\bibinfo {author} {\bibfnamefont {G.~M.}\ \bibnamefont
  {Sommers}}, \bibinfo {author} {\bibfnamefont {D.~A.}\ \bibnamefont {Huse}},\
  and\ \bibinfo {author} {\bibfnamefont {M.~J.}\ \bibnamefont {Gullans}},\
  }\href {https://doi.org/10.48550/ARXIV.2210.10808} {\bibinfo {title}
  {Crystalline quantum circuits}} (\bibinfo {year} {2022})\BibitemShut
  {NoStop}%
\bibitem [{\citenamefont {Li}\ \emph {et~al.}(2021)\citenamefont {Li},
  \citenamefont {Chen}, \citenamefont {Ludwig},\ and\ \citenamefont
  {Fisher}}]{Li2020}%
  \BibitemOpen
  \bibfield  {author} {\bibinfo {author} {\bibfnamefont {Y.}~\bibnamefont
  {Li}}, \bibinfo {author} {\bibfnamefont {X.}~\bibnamefont {Chen}}, \bibinfo
  {author} {\bibfnamefont {A.~W.~W.}\ \bibnamefont {Ludwig}},\ and\ \bibinfo
  {author} {\bibfnamefont {M.~P.~A.}\ \bibnamefont {Fisher}},\ }\bibfield
  {title} {\bibinfo {title} {{Conformal invariance and quantum nonlocality in
  critical hybrid circuits}},\ }\href
  {https://doi.org/10.1103/PhysRevB.104.104305} {\bibfield  {journal} {\bibinfo
   {journal} {Physical Review B}\ }\textbf {\bibinfo {volume} {104}},\ \bibinfo
  {pages} {104305} (\bibinfo {year} {2021})}\BibitemShut {NoStop}%
\bibitem [{\citenamefont {Zabalo}\ \emph {et~al.}(2022)\citenamefont {Zabalo},
  \citenamefont {Gullans}, \citenamefont {Wilson}, \citenamefont {Vasseur},
  \citenamefont {Ludwig}, \citenamefont {Gopalakrishnan}, \citenamefont
  {Huse},\ and\ \citenamefont {Pixley}}]{Zabalo2022}%
  \BibitemOpen
  \bibfield  {author} {\bibinfo {author} {\bibfnamefont {A.}~\bibnamefont
  {Zabalo}}, \bibinfo {author} {\bibfnamefont {M.~J.}\ \bibnamefont {Gullans}},
  \bibinfo {author} {\bibfnamefont {J.~H.}\ \bibnamefont {Wilson}}, \bibinfo
  {author} {\bibfnamefont {R.}~\bibnamefont {Vasseur}}, \bibinfo {author}
  {\bibfnamefont {A.~W.~W.}\ \bibnamefont {Ludwig}}, \bibinfo {author}
  {\bibfnamefont {S.}~\bibnamefont {Gopalakrishnan}}, \bibinfo {author}
  {\bibfnamefont {D.~A.}\ \bibnamefont {Huse}},\ and\ \bibinfo {author}
  {\bibfnamefont {J.~H.}\ \bibnamefont {Pixley}},\ }\bibfield  {title}
  {\bibinfo {title} {{Operator Scaling Dimensions and Multifractality at
  Measurement-Induced Transitions}},\ }\href
  {https://doi.org/10.1103/PhysRevLett.128.050602} {\bibfield  {journal}
  {\bibinfo  {journal} {Physical Review Letters}\ }\textbf {\bibinfo {volume}
  {128}},\ \bibinfo {pages} {050602} (\bibinfo {year} {2022})}\BibitemShut
  {NoStop}%
\bibitem [{\citenamefont {Ashcroft}\ and\ \citenamefont
  {Mermin}(1976)}]{Ashcroft1976}%
  \BibitemOpen
  \bibfield  {author} {\bibinfo {author} {\bibfnamefont {N.~W.}\ \bibnamefont
  {Ashcroft}}\ and\ \bibinfo {author} {\bibfnamefont {N.~D.}\ \bibnamefont
  {Mermin}},\ }\href@noop {} {\emph {\bibinfo {title} {{Solid State
  Physics}}}}\ (\bibinfo  {publisher} {Saunders College Publishing},\ \bibinfo
  {address} {New York},\ \bibinfo {year} {1976})\BibitemShut {NoStop}%
\end{thebibliography}
\end{document}